\documentclass{statsoc}

\usepackage[latin1]{inputenc}
\usepackage{graphicx}
\usepackage{amssymb, amsmath}
\usepackage{geometry}
\usepackage{rotating}

\newcommand{\ie}{\textit{i.e.}}
\newcommand{\fref}{Fig.~\ref}

\title{Spatial two tissue compartment model for DCE-MRI}

\author{Julia C.~Sommer}
\author[Julia C.~Sommer and Volker J.~Schmid]{Volker J.~Schmid}
\coaddress{Volker J.~Schmid, Department of Statistics, Ludwig-Maximilians-Universität, Ludwigstraße 33,
80539 München, Germany}
\email{volker.schmid@lmu.de}
\address{Ludwig-Maximilians-Universität, Munich, Germany.}

\begin{document}

\maketitle

\begin{abstract}
In the quantitative analysis of Dynamic Contrast-Enhanced Magnetic Resonance Imaging (DCE-MRI) compartment models allow to describe the uptake of contrast medium with biological meaningful kinetic parameters.
As simple models often fail to adequately describe the observed uptake behavior, more complex compartment models have been proposed.
However, the nonlinear regression problem arising from more complex compartment models often suffers from parameter redundancy.
In this paper, we incorporate spatial smoothness on the kinetic parameters of a two tissue compartment model by imposing Gaussian Markov random field priors on them.
We analyse to what extent this spatial regularisation helps to avoid parameter redundancy and to obtain stable parameter estimates. Choosing a full Bayesian approach, we obtain posteriors and point estimates running Markov Chain Monte Carlo simulations.
The proposed approach is evaluated for simulated concentration time curves as well as for \textit{in vivo} data from a breast cancer study.


\end{abstract}

\section{Introduction}

Nonlinear regression problems often suffer from parameter redundancy \citep{seber89}. A prominent example for nonlinear problems 
is compartment models, which are used in a variety of applications to model the exchange between different compartments \citep{schmid10,McNally2008,Eriksson1971,Herbst1963}. In this paper,
we will concentrate on redundancy issues in complex compartment models for Dynamic Contrast-Enhanced Magnetic Resonance Imaging (DCE-MRI).

DCE-MRI is an imaging technique which allows to image the perfusion in tissue \textit{in vivo}. After injection of a contrast medium, a 
series of images is obtained. This series shows the uptake dynamics of the contrast medium into the tissue over time. For example, in oncology, 
analysing the dynamics of the contrast medium allows to detect tumours, to specify malignancy and type of tumours, and to assess the success 
of cancer therapies \citep{padhani05,schmid2009mrm}. Typically, quantitative analysis of DCE-MRI is based on compartment models which capture the 
exchange of blood (containing contrast medium) between different, well-mixed compartments.
With the help of differential equations the form of the concentration time curve (CTC) can be analytically described depending on biologically meaningful 
kinetic parameters, resulting in a nonlinear regression problem.

\subsection{Compartment models}
Compartment models assuming various tissue architectures of different complexities have been proposed for quantitative analysis of DCE-MRI data.
The simplest and most frequently used models are the Tofts model, here also refered to as 1Comp model, and the "extended Tofts" model assuming only  the arterial plasma compartment and one interstitial space compartment~\citep{Tofts97,tofts99}.
In the Tofts model the observed CTC $C_t(t)$ is described by 
\begin{equation}
C_{t}(t) =   C_{p}(t) \ast K^{\text{trans}} \exp(-k_{\text{ep}}t), 
\end{equation} 
with $\ast$ the convolution operator, $C_p(t)$ the arterial input function (AIF), \textit{i.e.}, the concentration of contrast agent in the blood plasma, $K^{\text{trans}}$ the transfer rate from blood plasma to extracellular extravascular space (EES), and $k_{\text{ep}}$ the rate constant for transfer between EES and plasma.

Tumour tissue is often heterogeneous~\citep{schmid10} and the Tofts and extended Tofts models fail in describing its observed uptake dynamics~\citep{schmid09tmi}. 
Therefore, several authors propose more complex models to describe perfusion in tissue.   
For example, the two compartment exchange model (2CXM) has seperate compartments for arterial plasma and interstitial plasma~\citep{Brix2009,Sourbron2011}.
Multi-compartment models allow for two to three kinetically distinct tissue compartments to describe CTCs on a region of interest level~\citep{port99}.
Here, we use two tissue compartments (2Comp model) with the same architecture as \cite{port99} to model CTCs on a per voxel level:
\begin{equation}
C_{t}(t) =   C_{p}(t) \ast \left( K^{\text{trans}}_1 \exp(-k_{\text{ep}_1}t) + K^{\text{trans}}_2 \exp(-k_{\text{ep}_2}t) \right). 
\end{equation} 
With the 2Comp model on a voxel level CTCs in heterogeneous tissue can be more 
adequately described, especially at tumour margins \citep{Karcher2010} and tissue heterogeneity becomes accessible. This is important because tissue heterogeneity is diagnostically informative.
The 2Comp model is introduced in more detail in Section \ref{ssec:compartmentmodel} and compared to other compartment models in Section \ref{ssec:modelComparison}.

\subsection{Parameter redundancy}

Though necessary, increased model complexity brings challenges.
With increased complexity model parameters can become redundant and in this case cannot be stably estimated.

Parameter redundancy (or non-identifiability) is frequently encountered in nonlinear regression problems.
In contrast to identifiability problems occurring in standard linear regression models, problems occurring in nonlinear regression models are often such that they cannot be eliminated by optimal design \citep{Gilmour2012}, reparametrisation, nor reduced error variance.
This is the case when different nonlinear functions corresponding to different models or parameter constellations are too similar. 
Seber and Wild describe such an example for a sum of two or three exponentials \cite[p. 119]{seber89}:
\begin{quote}
\textit{Even though these functions are so different, the curves are visually indistinguishable. This approximate lack of identifiability is called parameter redundancy by Reich [1981]\nocite{Reich81}. Models such as this give rise to bad ill-conditioning, no matter where the x's are placed---a common problem with linear combinations of exponentials.}
\end{quote} 
We will discuss the redundancy problem for the 2Comp model in more detail in Section~\ref{redundancyexplained}.

As a solution to the redundancy issues we suggest to regularise the parameter space by spatially smoothing the parameter maps.
We propose a Bayesian framework in order to determine the kinetic parameters of the 2Comp model on a voxel level, using prior 
biological knowledge on the parameters and accounting for the spatial structure of the image. The approach makes use of the intrinsic
spatial information given by the voxel structure of the image. 
Spatial regularisation is done by Gaussian Markov random fields (GMRF) as priors on the kinetic parameters
as previously proposed by \cite{schmid06} and \cite{kelm2009} for the 1Comp model.

\subsection{Outline}
In this paper we aim to analyse to what extent the spatial regularisation helps to obtain stable parameter estimates in the 2Comp model.
We choose a full Bayesian approach and obtain posteriors and parameter estimates by running Markov Chain Monte Carlo (MCMC) simulations. 
The proposed model is evaluated for simulated CTCs and data from a breast cancer study.

An advantage of the Bayesian framework is that the posterior can still be computed in the case of parameter redundancy; however, the redundant parameters
will have multimodal marginal posteriors.
We find that assuming spatial smoothness on the exponential rates is an efficient way to regularise the parameter space and to make parameters 
of a 2Comp model identifiable.
As a result we obtain parameter estimates at a voxel level that are more stable. Hence, one can describe heterogeneity of the tissue without 
losing spatial information.

\section{Model}

The observed contrast concentration $Y_{i,j}$ at time $t_{j}$, $j=1,...,T$ in voxel $i=1,...,N$ can be described by the theoretical 
concentration time curve $C_t(t)$ depending on the voxel-specific kinetic parameters $\phi^{i}$ plus Gaussian noise, \textit{i.e.},
\begin{equation}
\label{equ:regression}
	 Y_{i,j} \sim N\left(C_{t}(\phi^{i}; t_{j}),\sigma_i^2\right),
\end{equation}
with $\sigma_i^2$ the voxel-specific variance of the Gaussian noise~\citep{schmid06}. 
The form of $C_{t}$ and the number of kinetic parameters is fixed, given a specific compartment model.

\subsection{Two tissue compartment model (2Comp)}
\label{ssec:compartmentmodel}
\begin{figure}[ht]
\centering	\includegraphics[scale=0.2]{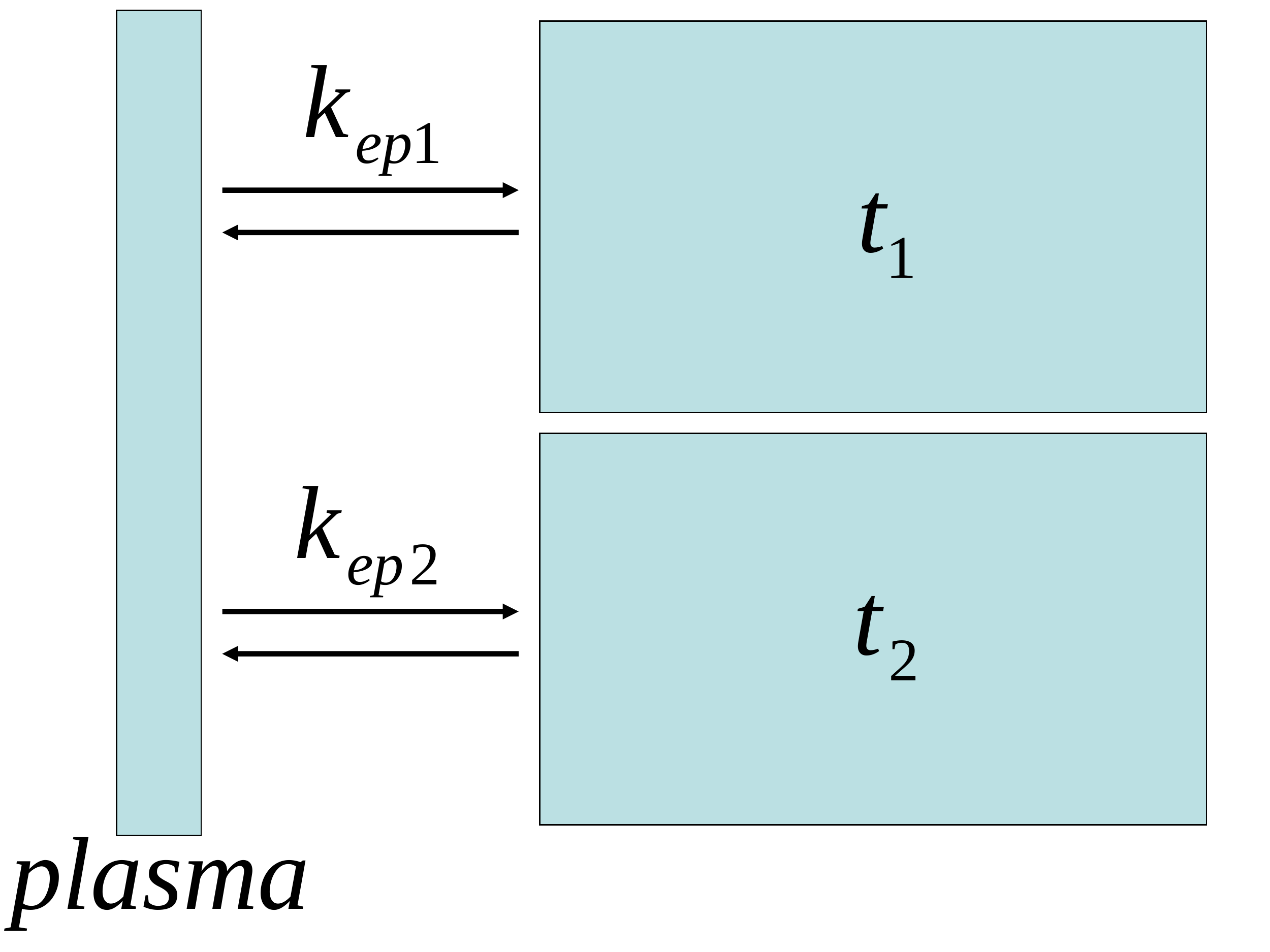}
\caption{Two tissue compartment model}
\label{Fig:2CompModel}
\end{figure}

Even at a voxel level, tissue can be heterogeneous. That is, there may be two tissue compartments with different kinetic properties
that exchange with plasma 
at distinct rates $k_{\text{ep}_{1}}$ and $k_{\text{ep}_{2}}$ (see Fig.~\ref{Fig:2CompModel}). Hence, we assume a model with two tissue compartments (2Comp).
We label the compartments such that $k_{\text{ep}_{1}} < k_{\text{ep}_{2}}$. This means that the first compartment is the slow and the second compartment is the fast exchanging compartment. 

As $k_{\text{ep}_{k}} = K^{\text{trans}}_k /v_{t_k}$~\citep{tofts99}, the volume $v_{t_k}$ of tissue $t_{k}$ per unit volume of tissue can be computed as $v_{t_k} = K^{\text{trans}}_k /k_{\text{ep}_{k}}$. The changes in tissue concentrations are given by
\begin{equation}
\begin{array}{lcr}
	v_{t_{1}} \frac{d}{dt}	C_{t_{1}}(t) = K^{\text{trans}}_{1} \left[C_{p}(t) - C_{t_{1}}(t)\right] \\
	v_{t_{2}} \frac{d}{dt}	C_{t_{2}}(t) = K^{\text{trans}}_{2} \left[C_{p}(t) - C_{t_{2}}(t)\right].
\end{array}
\end{equation}
The solution of these differential equations is given by
\begin{equation}
\label{equ:Ctj}
C_{t_{k}}(t) = C_{p}(t)  \ast \frac{K^{\text{trans}}_{{k}}}{v_{t_{k}}} exp\left(-\frac{K^{\text{trans}}_{{k}}}{v_{t_{k}}}t\right)
\end{equation}
for k=1,2 denoting the different tissue compartments. The total (observable) concentration is then given as $C_{t} = v_{t_{1}}C_{t_{1}} + v_{t_{2}}C_{t_{2}}$ by
\begin{equation}
\label{equ:Ct} 
C_{t}(t) = \sum^{2}_{k=1} C_{p}(t) \ast K^{\text{trans}}_{{k}} exp(-k_{\text{ep}_{k}}t) .
\end{equation}
 
The AIF describes the input of contrast agent through the blood stream. As suggested by \cite{tofts91}, we use a bi-exponential function of the form
\begin{equation}
		C_{p}(t) = D \sum^{2}_{l=1} a_{l} exp(-m_{l}t)
\end{equation}
with dose $D$ and values $a_{1} = 3.99$ $kg/l$, $a_{2} = 4.78$ $kg/l$, $m_{1} = 0.144$ $min^{-1}$, $m_{2} = 0.0111$ $min^{-1}$. 

We use an exponential parametrisation that insures the rate and transfer constants to be positive \citep[see][and references therein]{schmid06}:
		 $\theta_{k}^{i}  = \log \left(  k^{i}_{\text{ep}_{k}}\right)$, 
		$\gamma_{k}^{i}   = \log\left( K^{\text{trans},i}_{k} \right)$, for $k=1,2$.

\subsection{Relation to other compartment models}
\label{ssec:modelComparison}
In the proposed 2Comp model, the observed concentration $C_{t}(t)$ is described by an impulse response function (sum of two exponentials) convolved with the AIF, see equation (\ref{equ:Ct}). 
In the 2CXM the interstitial space and the interstitial plasma are modeled with seperate compartments.
Though explained by different compartment designs, the impulse response of the 2CXM is also a sum of two exponentials. 
Hence, the 2Comp model and the 2CXM lead to the same nonlinear regression problem.
We prefer to use the 2Comp model due to the charming fact that the impulse response is directly expressed by interpretable parameters  $K^{\text{trans}}_{{1}}$, $K^{\text{trans}}_{{2}}$, $k_{\text{ep}_{1}}$ and $k_{\text{ep}_{2}}$.
In contrast, in the 2CXM the impulse response is expressed by auxiliary variables (called $F_{+}$, $F_{-}$, $K_{+}$ and $K_{-}$) which are complicated functions of interpretable quantities---see Lemma 3 of~\cite{Sourbron2011}.

For the case that the exchange rates are the same, $k_{\text{ep}_{1}}=k_{\text{ep}_{2}}$, or when one of the tissue volumes vanishes, $v_{t_{1}}=0$ or $v_{t_{2}}=0$, the impulse response reduces to a single exponential and the 2Comp model corresponds to the standard Tofts model, also referred to as 1Comp model here.

For the case that one exchange rate becomes infinite,  $k_{\text{ep}_{2}}=\infty$, the observed concentration is of the form
\begin{equation}
\label{equ:CtETM}
C_{t}(t) = v_{t_{2}} C_{p}(t) +  C_{p}(t) \ast K^{\text{trans}}_{{1}} exp(-k_{\text{ep}_{1}}t). 
\end{equation}
In this case the second tissue compartment takes the role of an interstitial plasma compartment, $C_{t_{2}}=C_{p}$,
and the 2Comp model corresponds to the extended Tofts model.

\subsection{Redundancy issues in the independent voxelwise model}
\label{sec:voxelwiseModel}
In a voxelwise approach, the CTCs of all voxels are fitted independently from each other. 
Similar to the voxelwise Bayesian 1Comp model evaluated in \cite{schmid06}, 
we impose Gaussian priors on the logarithmic rate constants $\theta_{k}^{i}$ 
$$
	\theta_{k}^{i}  | \tau_{\theta_{k}} \sim N(\mu_{\theta_{k}}, (\tau_{\theta_{k}})^{-1}) 
$$
and on the logarithmic transfer constants $\gamma_{k}^{i}$
$$
	\gamma_{k}^{i} | \tau_{\gamma_{k}} \sim N(\mu_{\gamma_{k}}, (\tau_{\gamma_{k}})^{-1})
$$
independently for all voxels $i=1,\ldots,N$
with fixed precisions $\tau_{\theta_{k}}= \tau_{\gamma_{k}} =1$ and
$\mu_{\theta_{1}}=\mu_{\gamma_{1}}=\mu_{\gamma_{2}}=0$, $\mu_{\theta_{2}}=\log(5)$ .
With this prior, all rate and transfer constants $k^{i}_{\text{ep}_{k}}$ and $K^{\text{trans},i}_{k}$ remain positive. Rate and transfer constants of the first compartment with \textit{a priori} probability of $99.86\%$ do not exceed 20 $min^{-1}$. The dynamics in the second compartment is assumed to be faster with \textit{a priori} expected $k^{i}_{\text{ep}_{2}}$ values of five.

\begin{figure}[ht]
\centering
\includegraphics[scale=0.5]{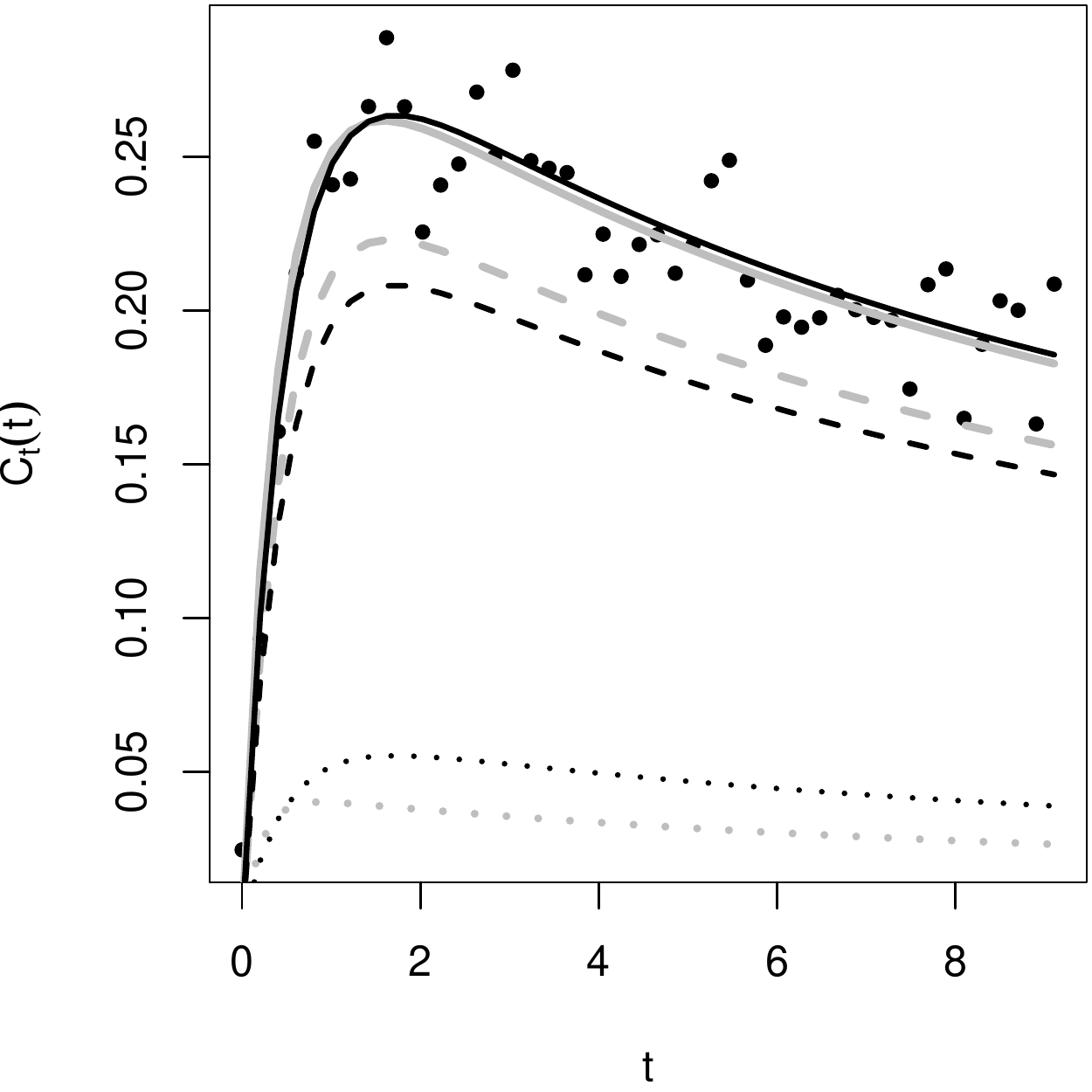}
\caption{Similar CTCs for two different parameter vectors. 
Black line: CTC described by $k_{\text{ep}_1}=2.07$, $k_{\text{ep}_2}=2.07$, $K_1^{\text{trans}}=0.55$, $K_2^{\text{trans}}=0.15$ (can as well be described by only one compartment with $k_{\text{ep}_1}=2.07$ and $K_1^{\text{trans}}=0.7$).
Grey line: CTC described by $k_{\text{ep}_1}=2.19$, $k_{\text{ep}_2}=5.02$, $K_1^{\text{trans}}=0.62$, $K_2^{\text{trans}}=0.24$.
The dashed lines show the contribution of the first compartment and the dotted lines those of the second compartment.
}
\label{Fig:redundantfit}
\end{figure}

\label{redundancyexplained}
This independent voxelwise model leads to unstable estimates due to redundancy issues~\citep{Karcher2010}. 
Obviously, redundancy issues occur when the contribution of one compartment vanishes. 
However, they may as well occur when the exponential rates are too similar.
There are theoretical results on parameter redundancy in sum of exponentials models \citep{seber89}, however, a generalisation for
the case of convolved exponentials is tricky. 
In \cite{Reich81} a redundancy measure was used to show that parameters in a sum of two exponentials model are 
highly redundant if the exponential rates differ by less than a factor of five. 
Even though this result does not directly transfer to the convolved exponentials given in equation (\ref{equ:Ct}),
this results still helps to understand parameter redundancy in the 2Comp model. 
In Fig.~\ref{Fig:redundantfit} we show an example for data simulated from a two-compartment model where the exponential rates
differ by a factor of four.
In this case, a solution from a compartment model with only one tissue compartment exists that fits the observed concentration 
reasonably well. Therefore, several quite distinct parameter vectors describe very similar CTCs, meaning that parameters are 
redundant. 

\subsection{Spatial regularisation}
In order to solve the problem of redundancy, we propose to use spatial regularisation. In the following, we will
introduce a spatial prior which accounts for the spatial information intrinsic in the DCE-MR images. 

In the proposed spatial model, we assume that rate constants $k^{i}_{\text{ep}_{k}}$ vary smoothly in space and hence that the exchange properties of each tissue compartment are rather smooth.
In contrast, the contribution of differently behaving compartments in each voxel is assumed to be quite flexible, meaning that the tissue volumes $v_{t_k}^{i} = K^{\text{trans},i}_k/k_{\text{ep}_{k}}^{i}$ may vary unsmoothly from voxel to voxel.
Then the transfer rate as a product of rate and volume $ K^{\text{trans},i}_k = k_{\text{ep}_{k}}^{i} v_{t_k}^{i}$ inherits the spatial smoothness of $k_{\text{ep}_{k}}^{i}$, but is less smooth due to varying $v_{t_k}^{i}$ values.
The spatial smoothness of the kinetic parameters is modelled using a Gaussian Markov random field on its logarithmic transforms $\theta_{k}^{i}$, $\gamma_{k}^{i}$ \citep{rueheld2005, schmid06}.

We use a neighbourhood structure where adjacent voxels are neighbours, that is, each voxel has four neighbours unless it lies at the edge of the image.
From this, a prior distribution can be defined by assuming a Gaussian distribution on the differences of neighbouring logarithmic rate and transfer constants:
$$
	\theta_{k}^{i} - \theta_{k}^{j} | \tau_{\theta_{k}} \sim N(0, (\tau_{\theta_{k}})^{-1}) \mbox{ for }  i \sim j 
$$
and 
$$
	\gamma_{k}^{i} - \gamma_{k}^{j} | \tau_{\gamma_{k}} \sim N(0, (\tau_{\gamma_{k}})^{-1}) \mbox{ for }  i \sim j.
$$

This spatial prior on the logarithmic rates leads to smooth parameter maps of $k_{\text{ep}_1}^{i}$ and $k_{\text{ep}_2}^{i}$,  $K_1^{\text{trans},i}$ and $K_2^{\text{trans},i}$.
However, \textit{a priori} we expect much smoother maps for $k_{\text{ep}_1}^{i}$ and $k_{\text{ep}_2}^{i}$ and less smooth maps for $K_1^{\text{trans},i}$ and $K_2^{\text{trans},i}$.
Hence, we use Gamma priors on the precisions
$\tau_{\theta_{k}} \sim Ga(a_{\theta_{k}},b_{\theta_{k}})$ for $k=1,2$ with $a_{\theta_{1}}=a_{\theta_{2}}=1000$ and $b_{\theta_{1}}=b_{\theta_{2}}=1$
and
$\tau_{\gamma_{k}} \sim Ga(a_{\gamma_{k}},b_{\gamma_{k}})$ for $k=1,2$  with $a_{\gamma_{1}}=a_{\gamma_{2}}=0.0001$ and $b_{\gamma_{1}}=b_{\gamma_{2}}=0.01$.

Furthermore, we assume the noise variance to be the same $\sigma_{i}^{2} = \sigma^{2}$ for all voxels $i$.
For the observation variance we assume an Inverse Gamma prior
$\sigma_{i}^2 \sim IG(a,b)$
with $a$ and $b$ such that the \textit{a priori} expected SNR corresponds to values typically observed in breast tumour DCE-MRI data (SNR typically ranges from  10 to 20).
We choose this prior to be more informative with increasing number of voxels.

\subsection{Implementation}
We implemented the proposed spatial 2Comp model extending the R-package \texttt{"dcemriS4"} \citep{whitcher09,whitcher11}.
For each voxel, we simulate from the posterior of the model parameters with a MCMC algorithm \citep{gilks96}.
Starting with random values, the voxels are subsequently updated in random order.
More precisely, starting values are drawn from uniform distributions per voxel with $v^{start}_{t_{1}} \sim U[0,1]$, $v^{start}_{t_{2}}=1- v^{start}_{t_{1}}$, and $k^{start}_{\text{ep}_1} \sim  U[0.1,0.3]$, $k^{start}_{\text{ep}_2} \sim  U[1.75,5.25]$.
The logarithmic rate and transfer constants $\theta_{1}^{i}$, $\theta_{2}^{i}$, $\gamma_{1}^{i}$, $\gamma_{2}^{i}$ are updated 
with Metropolis Hastings steps with random walk proposal.
Gibbs update steps are used for the hyperparameters $1/\sigma^2$, $\tau_{\theta_{k}}^{i}$ and $\tau_{\gamma_{k}}^{i}$,
as its full conditionals are Gamma distributions that can be sampled from directly, see Appendix \ref{sec:appendixfc}.

The proposal variances of the random walk proposals are tuned such that
the Metropolis-Hastings acceptance rates are approximately 20\%.
After a burn-in of 5,000 iterations, 5,000 iterations are performed with every third sample saved. For parameter point estimation we use the median of the MCMC sample.

\subsection{Measure of model complexity}
\label{ssec:Modelfit}
We suggest to use the number of effective parameters $p_{D}$ as a measure of model complexity and, hence, tissue heterogeneity. 
The number of effective parameters $p_{D}$ is calculated as the difference of the posterior median of the deviance and the deviance evaluated at the posterior median value~\citep{spiegelhalter02}. It was introduced for the calculation of deviance information criterion (DIC) and is typically used to asses model fit and complexity in Bayesian hierarchical models. The DIC is defined as the posterior median deviance plus the number of effective parameters $p_{D}$~\citep{spiegelhalter02}.

However, as \cite{spiegelhalter02} point out, $p_{D}$ can become negative in cases where the posterior mean or posterior median is a poor estimator. This is certainly the case when dealing with multimodal posteriors due to parameter redundancy. Hence, we can detect redundancy issues by looking at the $p_D$.

Here, although we are dealing with a joint model for all voxels, \ie, a model for the whole image, we will also compute a voxelwise $p_D$ using the deviance in each voxel. The deviance is $ D(\phi^{i},\sigma^2) = -2*l(\phi^{i},\sigma^2)$ where $l(\phi^{i},\sigma^2)$ is the log-likelihood function
given in Appendix \ref{sec:appendixfc}. 
It is evaluated at the posterior median values of $\phi^{i}$ and $\sigma^2$ in order to calculate the deviance of the median and
it is evaluated at each sample value of $\phi^{i}$ and $\sigma^2$ in order to calculate the median deviance.
This allows to assess the model complexity per voxel and, hence, the tissue heterogeneity.  

\section{Simulation study}
\subsection{Simulation setup}
\begin{figure}[t]
\centering	\includegraphics[scale=0.3]{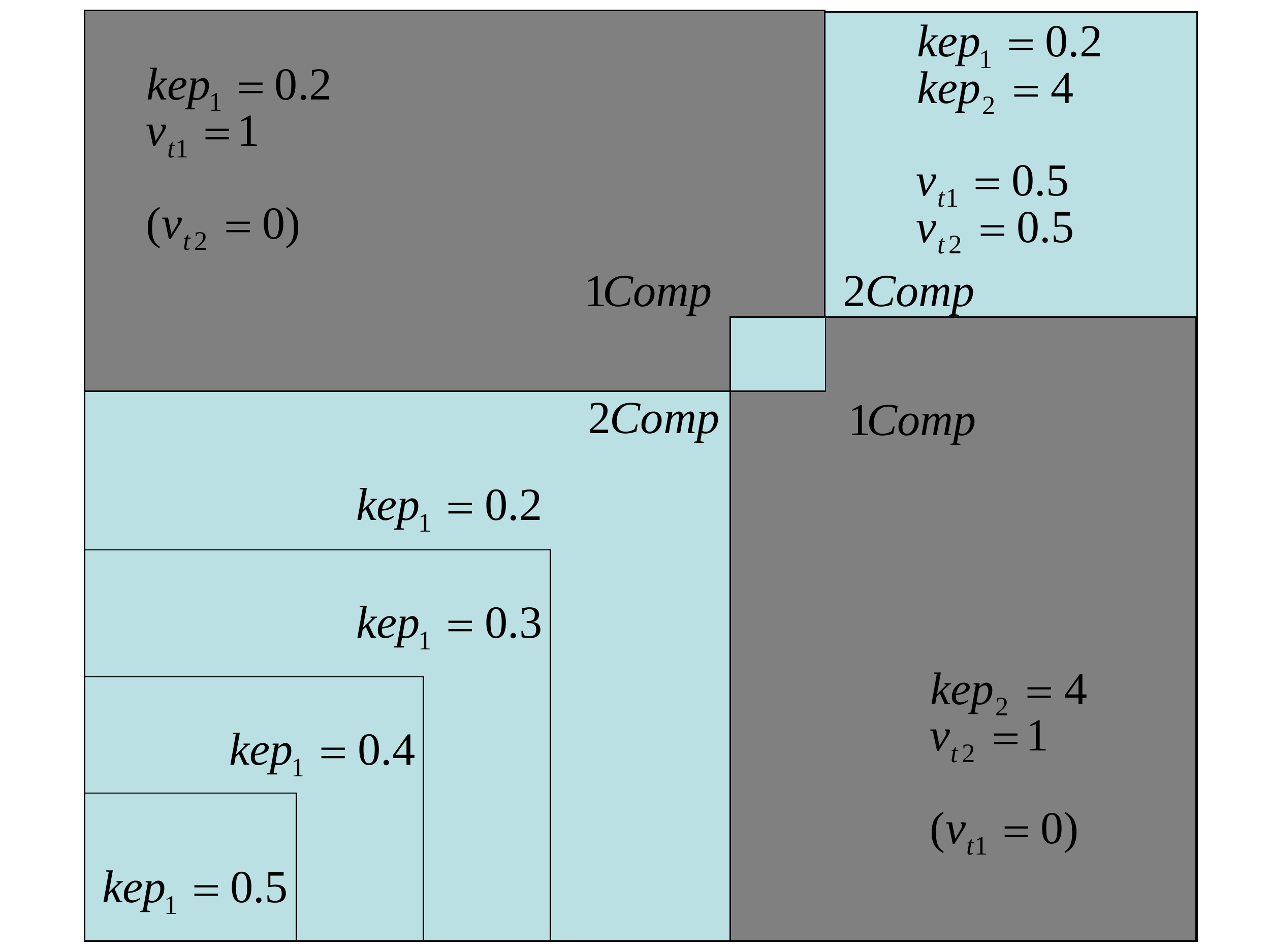}
\caption{Sketch of simulation design}
\label{Fig:simulationDesign}
\end{figure}
In order to evaluate the proposed voxelwise and spatial 2Comp models, we simulated a DCE-MR image of $25 \times 25$ voxels 
with different parameter combinations in a two tissue compartment model. The parameter configuration is given in Fig.~\ref{Fig:simulationDesign}.
For three blocks of different size, we simulated CTCs from a true 2Comp model, \ie, a mixture of two tissue compartments with very different exchange rates $k_{\text{ep}_1} = 0.2$ and $k_{\text{ep}_2}=4$. Both compartments contribute equally with volumes $v_{t_{1}} = v_{t_{2}} = 0.5$. In the lower left block, the exchange rate $k_{\text{ep}_1}$ varied smoothly from $0.2$ in the middle to $0.5$ at the corner.

For two blocks, we simulated CTCs from a 1Comp model. One of those blocks is described by a tissue compartment exchanging rather slowly with plasma at rate $k_{\text{ep}_1} = 0.2$ and having volume $v_{t_{1}} =1$  (the fast exchanging compartment with $k_{\text{ep}_2}=4$ has no contribution, i.e.\ $v_{t_{2}}=0$).
For the other block, the exchange with plasma is rather fast: $k_{\text{ep}_2}=4$, $v_{t_{2}}=1$  (the slow exchanging compartment $k_{\text{ep}_1} = 0.2$ has no contribution, i.e.\ $v_{t_{1}}=0$).

Within each block, uniformly distributed noise $U[0.8,1.2]$ was multiplied to the parameters $k_{\text{ep}_1}$, $k_{\text{ep}_2}$, $K_1^{\text{trans}}$ and $K_2^{\text{trans}}$ per voxel and the corresponding CTC was computed from these kinetic parameters.
Gaussian noise was added to the simulated CTCs with standard deviation $\sigma = 0.05$.

\subsection{Results}
\label{sec:simnonspat}

\begin{figure}[t]
\begin{tabular}{cc}
\includegraphics[scale=0.4]{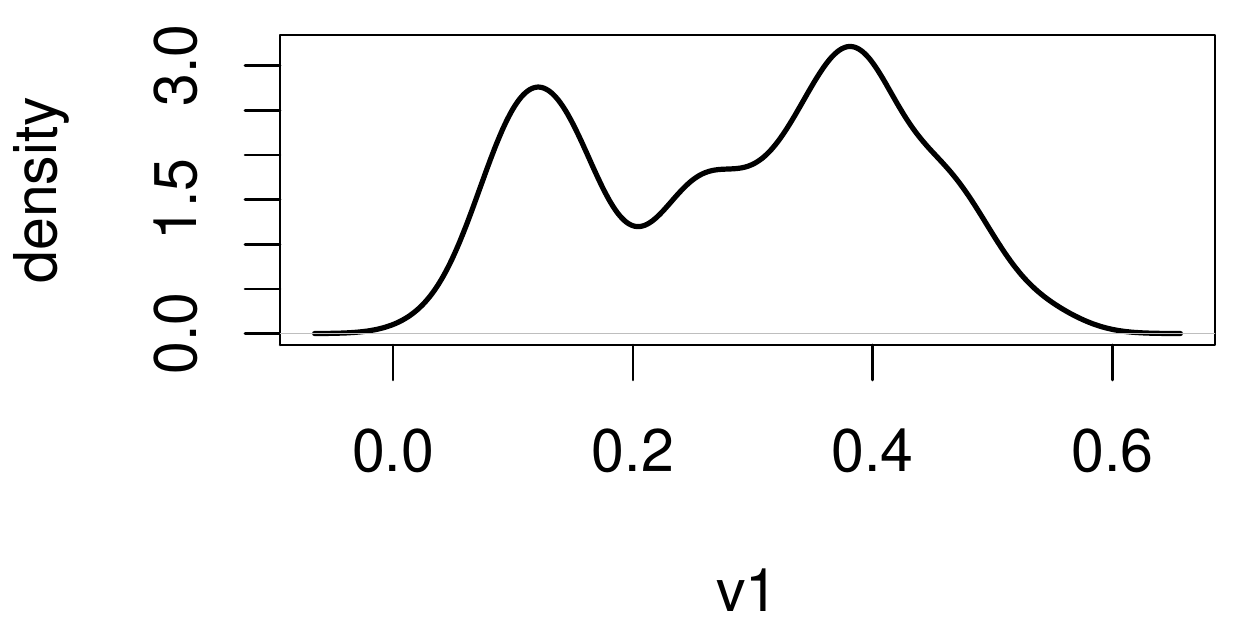}&
\includegraphics[scale=0.4]{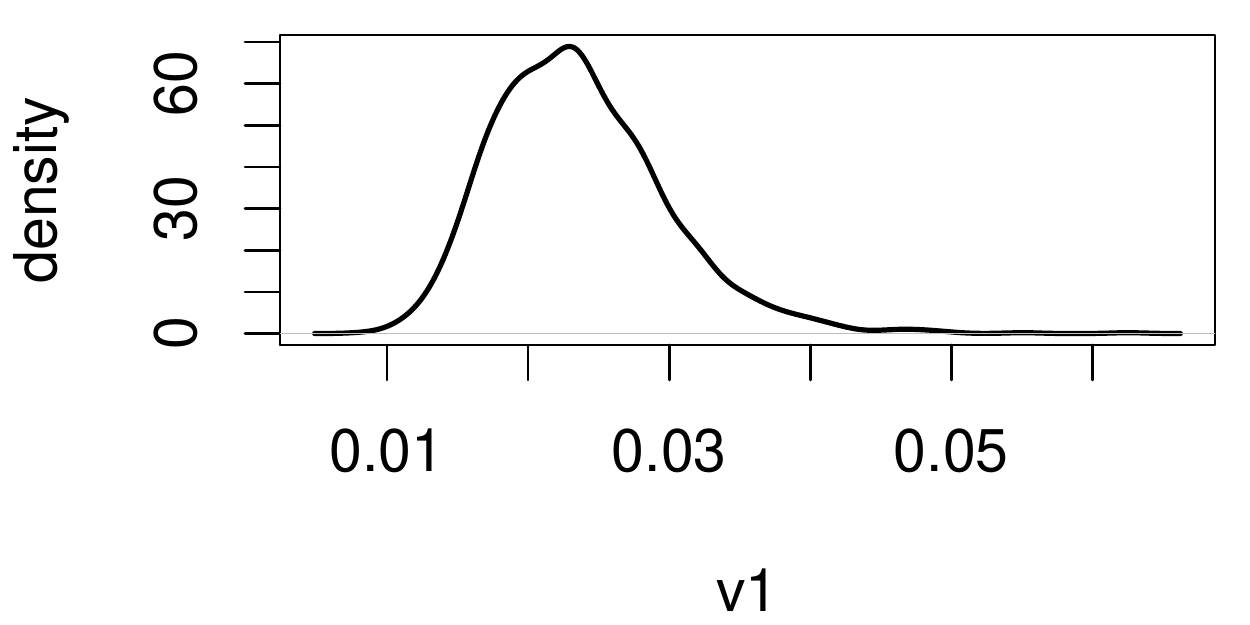}\\
\includegraphics[scale=0.4]{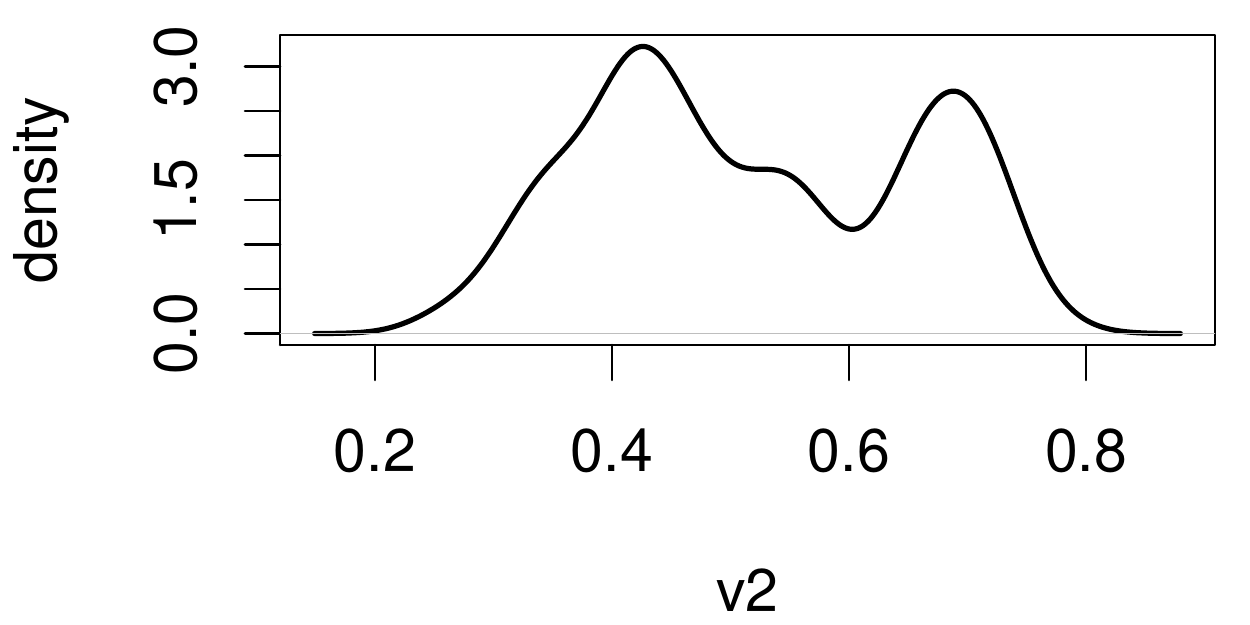}&
\includegraphics[scale=0.4]{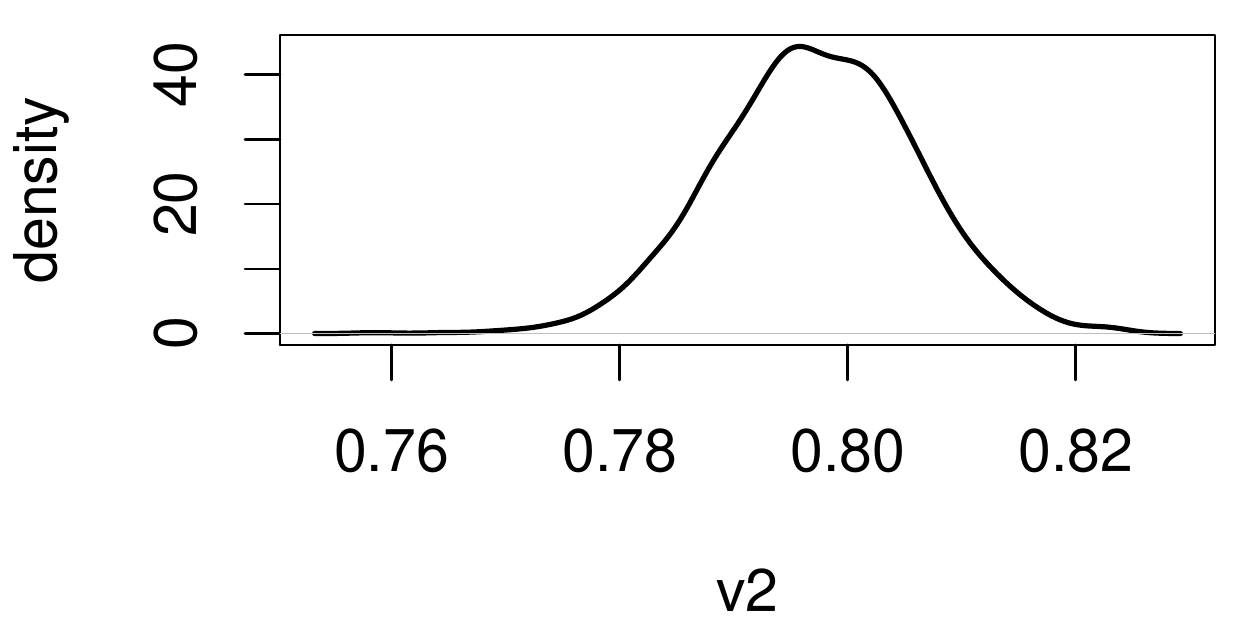}\\
\includegraphics[scale=0.4]{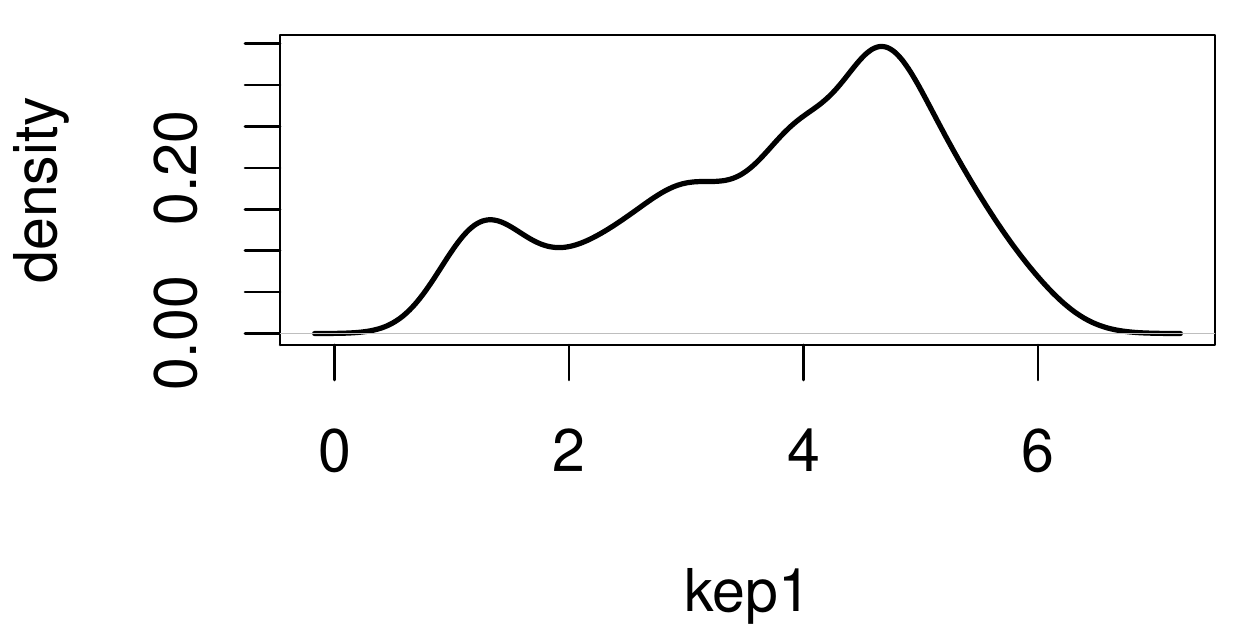}&
\includegraphics[scale=0.4]{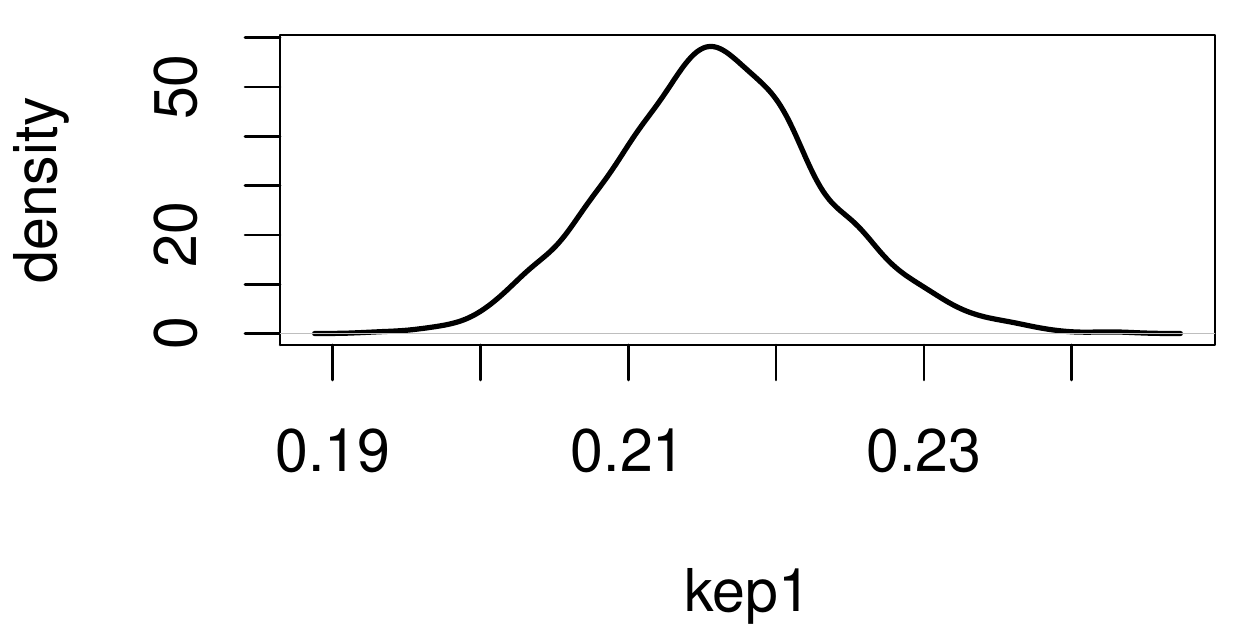}\\
\includegraphics[scale=0.4]{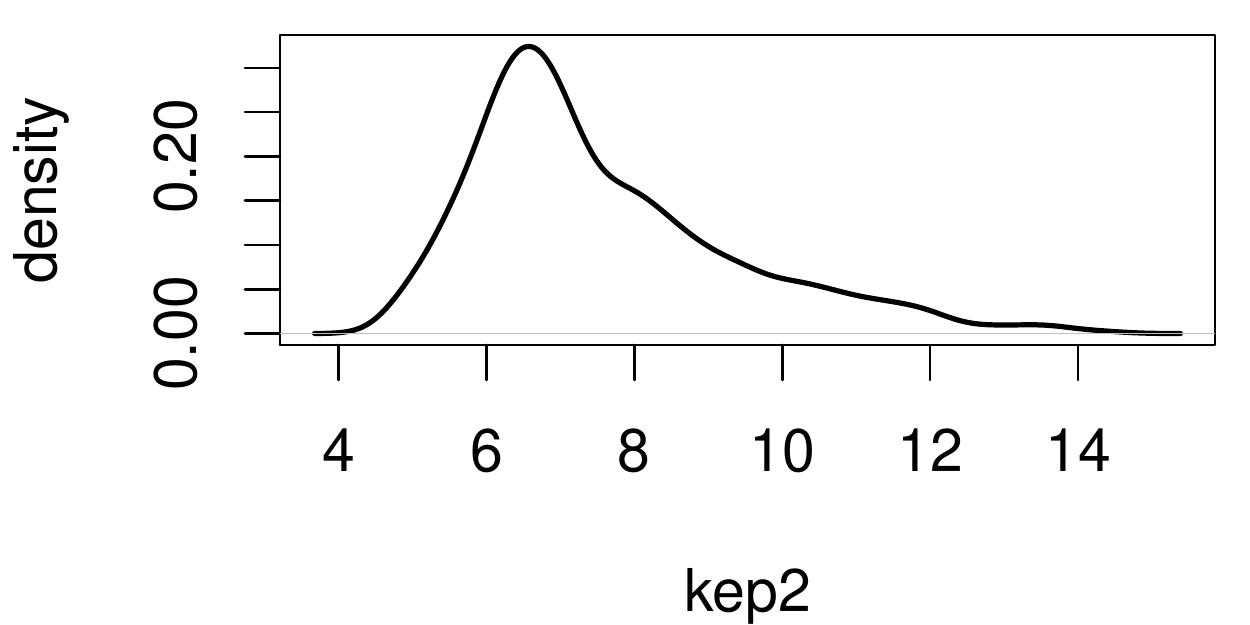}&
\includegraphics[scale=0.4]{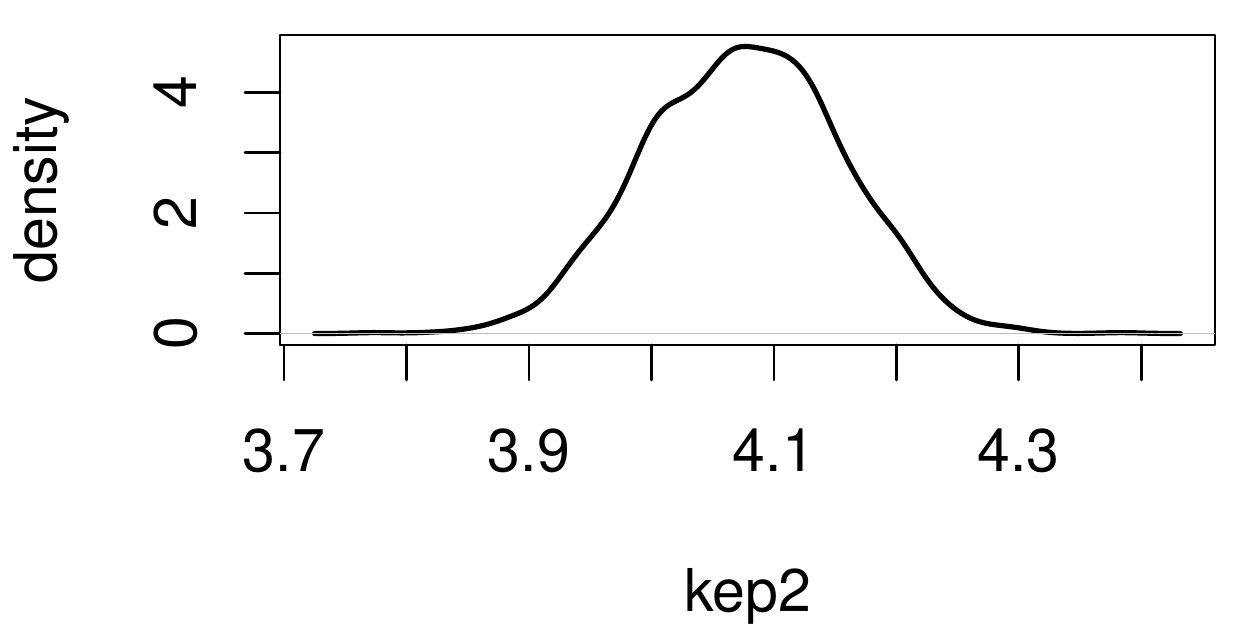}
\end{tabular}
\caption{Posterior marginal densities for curve simulated from 1Comp.  
Left: Voxelwise 2Comp fitted. Right: Spatial 2Comp fitted}
\label{Fig:posteriors}
\end{figure}
As discussed above, when fitting a model with two tissue compartments one often deals with identifiability issues. 
In these cases, the model is overparametrised and one observes unstable estimates of all parameters.
The Bayesian approach allows to evaluate the posterior anyway; however, in the case of redundancy, the marginal posteriors typically
are multimodal.

For example, in the blocks simulated from a 1Comp model parameter estimates are redundant and hence unstable in the lower right of the simulated image.
For one of these voxels, Fig.~\ref{Fig:posteriors} depicts the marginal posteriors of volume fractions $v_{t_k}$ and rate constants $k_{\text{ep}_k}$.
With the voxelwise approach, the posteriors are multimodal and hence there is no good point estimator.
In comparison, the spatial approach
produces unimodal posteriors and good point estimates can be gained by computing the median of the MCMC sample. In contrast to the voxelwise model, the contribution of the two compartments are well separated ($k_{\text{ep}_1}$ and $k_{\text{ep}_2}$ samples are not too similar) and the estimated volume of the first compartment is close to zero.

\begin{figure}[t]
\centering
\includegraphics[scale=0.45]{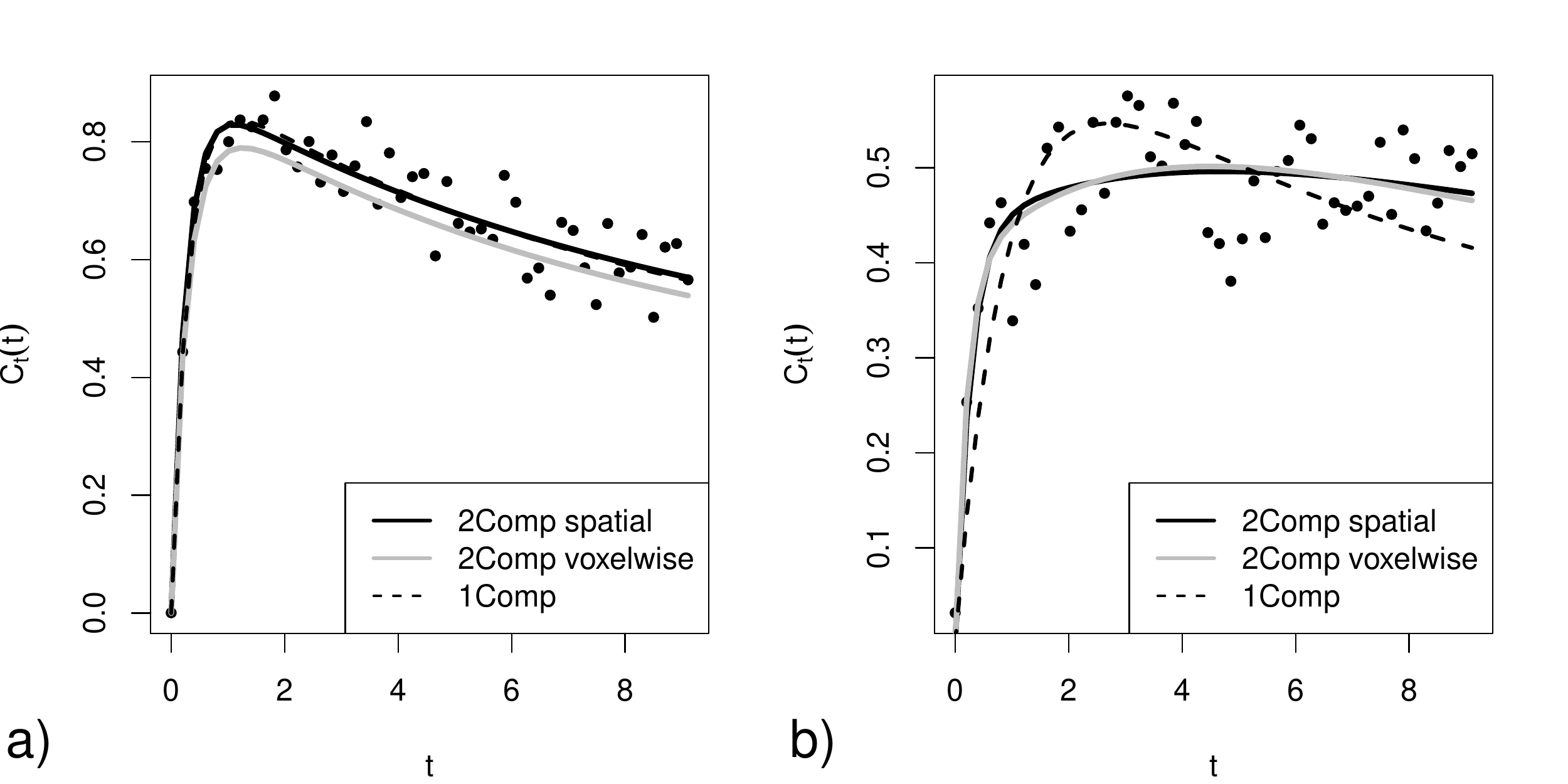}
\caption{Simulated data and curve fits for 1Comp model, spatial and voxelwise 2Comp models. 
(a) Data simulated from a 1Comp model (parameter constellation as in the lower right corner).
(b) Data simulated from a 2Comp model (parameter constellation as in the upper right corner). }
\label{Fig:simcurves}
\end{figure}

Using the point estimates for the kinetic parameters we obtain estimated CTCs and refer to them as fit.
In \fref{Fig:simcurves} we compare the fit of the 1Comp model, the voxelwise, and the spatial 2Comp model. 
For a curve simulated from a 1Comp model (\fref{Fig:simcurves} (a)), the fit of the spatial 2Comp model is similar to the fit of 
the 1Comp model. 
However, the voxelwise 2Comp model fails to adequately fit the curve due to redundancy issues. 
The SSE is about 0.14 for the voxelwise 2Comp model and about 0.1 for the 1Comp model as well as the spatial 2Comp model.
For a curve simulated from a 2Comp model (\fref{Fig:simcurves} (b)), both the spatial and voxelwise 2Comp 
models clearly outperform the 1Comp model with similar fits.
Here, the SSE is about 0.12 for the spatial and voxelwise 2Comp models and 0.2 for the 1Comp model.

\begin{figure}[t]
\begin{tabular}{ccc}
& 1Comp \hspace{2cm} 2Comp \hspace{1.8cm} 2Comp &\\
& \hspace{3.0cm} voxelwise \hspace{1.8cm} spatial &\\
\begin{rotate}{90}\hspace{10mm}SSE\end{rotate}&
\includegraphics[scale=0.2]{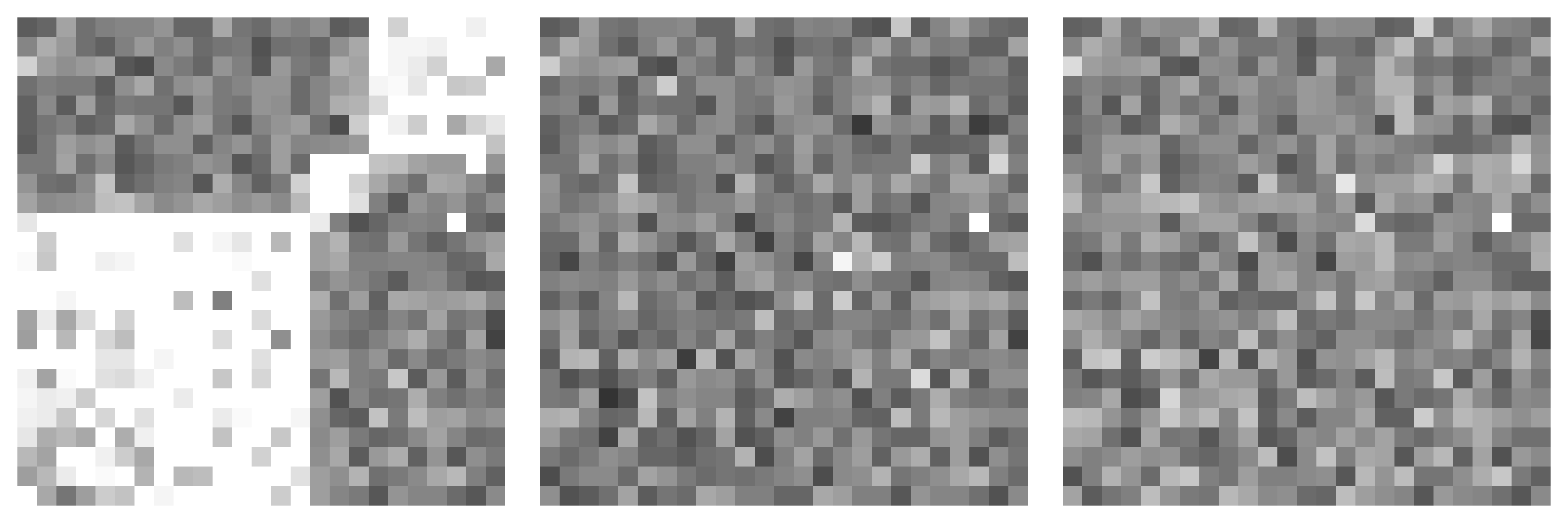}&
\includegraphics[scale=0.3]{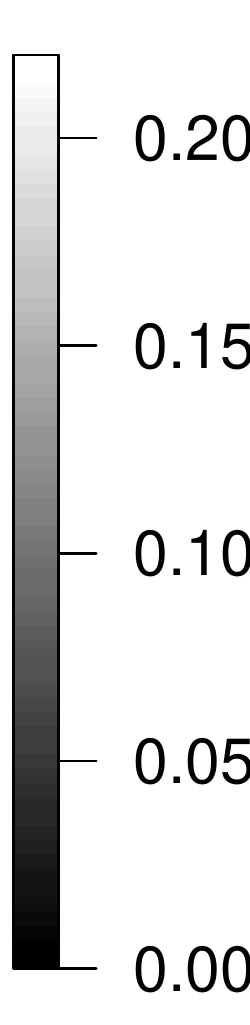}\\
\begin{rotate}{90}\hspace{10mm}$p_D$\end{rotate}&
\includegraphics[scale=0.2]{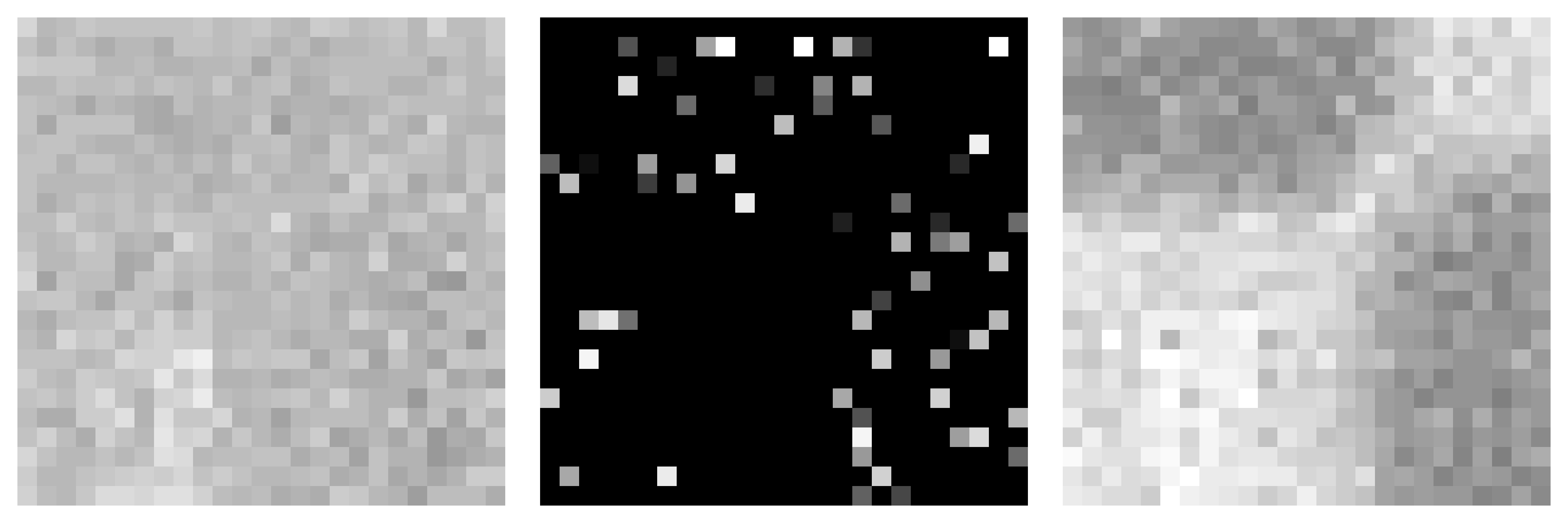}&
\includegraphics[scale=0.3]{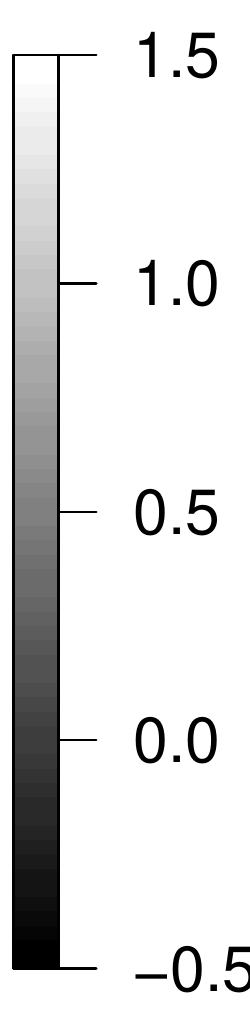}
\end{tabular}
\caption{Evaluation of model fit: sum of squared error (SSE) and $p_D$}
\label{Fig:mapsSimulationFit}
\end{figure}

In Fig.~\ref{Fig:mapsSimulationFit} the sum of squared errors (SSE) per voxel are depicted for the 1Comp and  
the voxelwise and the spatial 2Comp model. 
Considerable differences in SSE for the 1Comp model compared to both 2Comp models can be observed for the three blocks simulated from a true 2Comp model.
The voxelwise and spatial 2Comp models have similar SSE, with increased SSE in the voxelwise model for voxels with multimodal posteriors. These differences cannot be distinguished at this scale and were shown for a specific curve above (\fref{Fig:simcurves} (a)).

For voxels with multimodal posteriors, the estimates of $p_D$ are not meaningful. In the voxelwise 2Comp model estimated $p_D$ values are often negative due to parameter redundancy. In contrast, $p_D$ values are always positive in the spatial 2Comp model as the posteriors are unimodal. Values  range between 0.5 and 1 with median 0.68 in the 1Comp blocks. In the 2Comp blocks, $p_D$ values between 0.8 and 1.6 with median 1.2 show increased tissue heterogeneity.

\begin{figure}[ht]
\begin{tabular}{ccc}
&\hspace{-1.0cm}voxelwise \hspace{2.0cm} spatial \hspace{2.0cm} true&\\
\begin{rotate}{90}\hspace{10mm}$k_{\text{ep}_1}$\end{rotate}&
\includegraphics[scale=0.2]{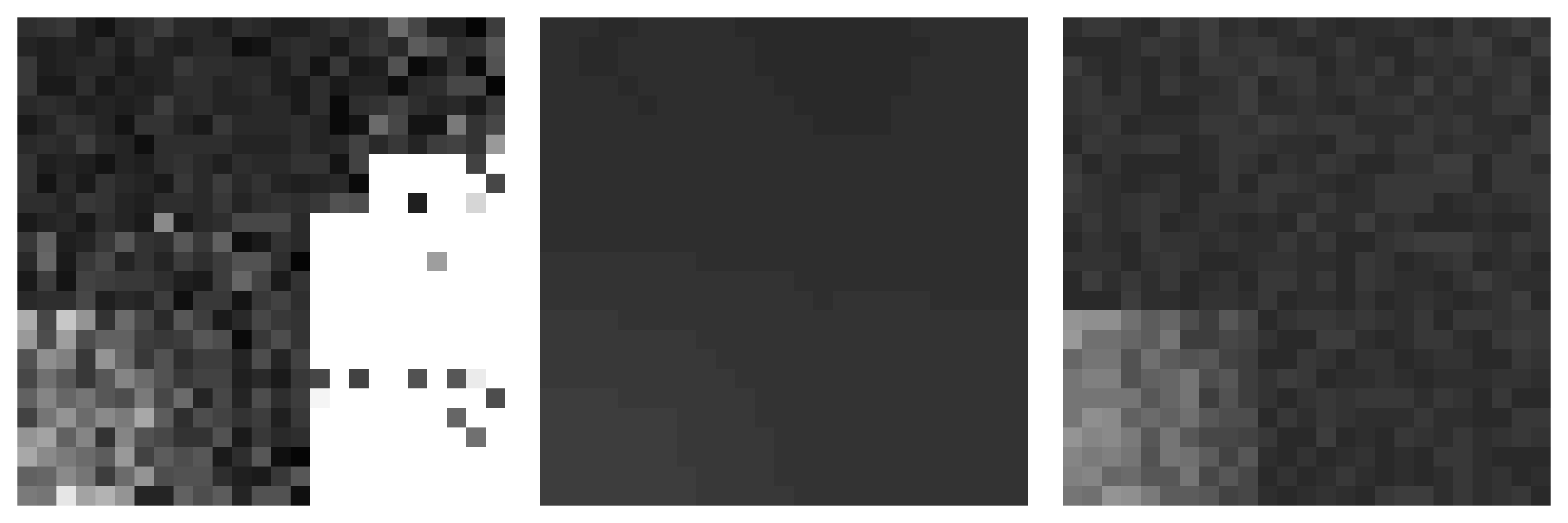}&
\includegraphics[scale=0.3]{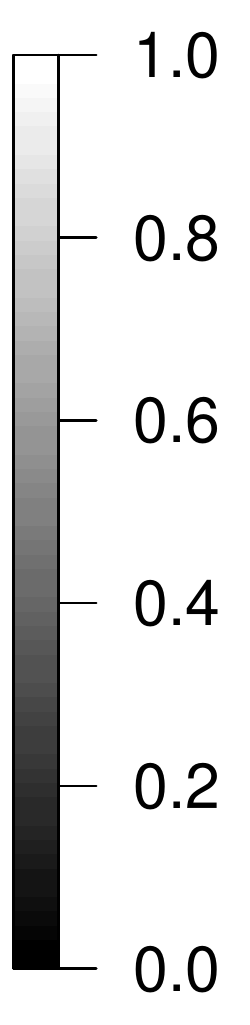}\\
\begin{rotate}{90}\hspace{10mm}$k_{\text{ep}_2}$\end{rotate}&
\includegraphics[scale=0.2]{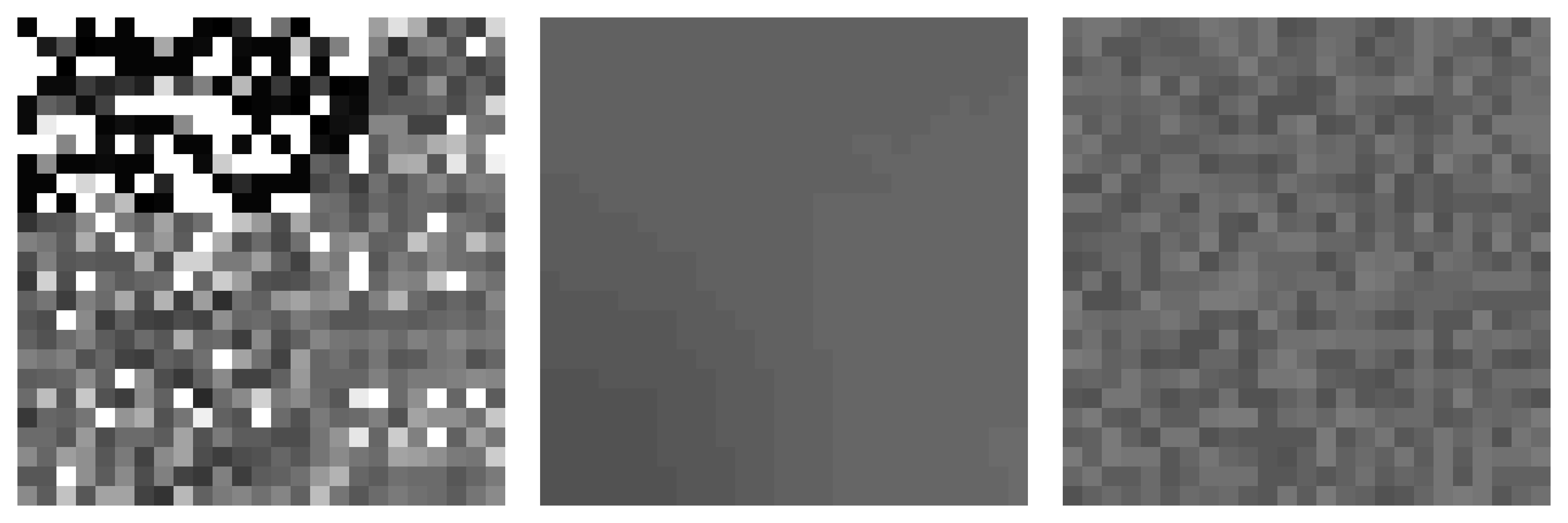}&
\includegraphics[scale=0.3]{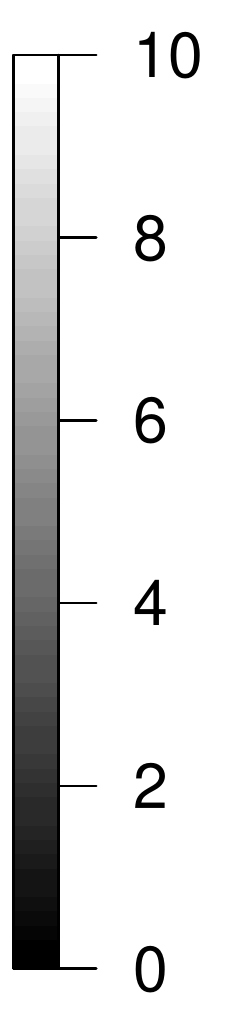}\\
\begin{rotate}{90}\hspace{10mm}$K^{\text{trans}}_{1}$\end{rotate}&
\includegraphics[scale=0.2]{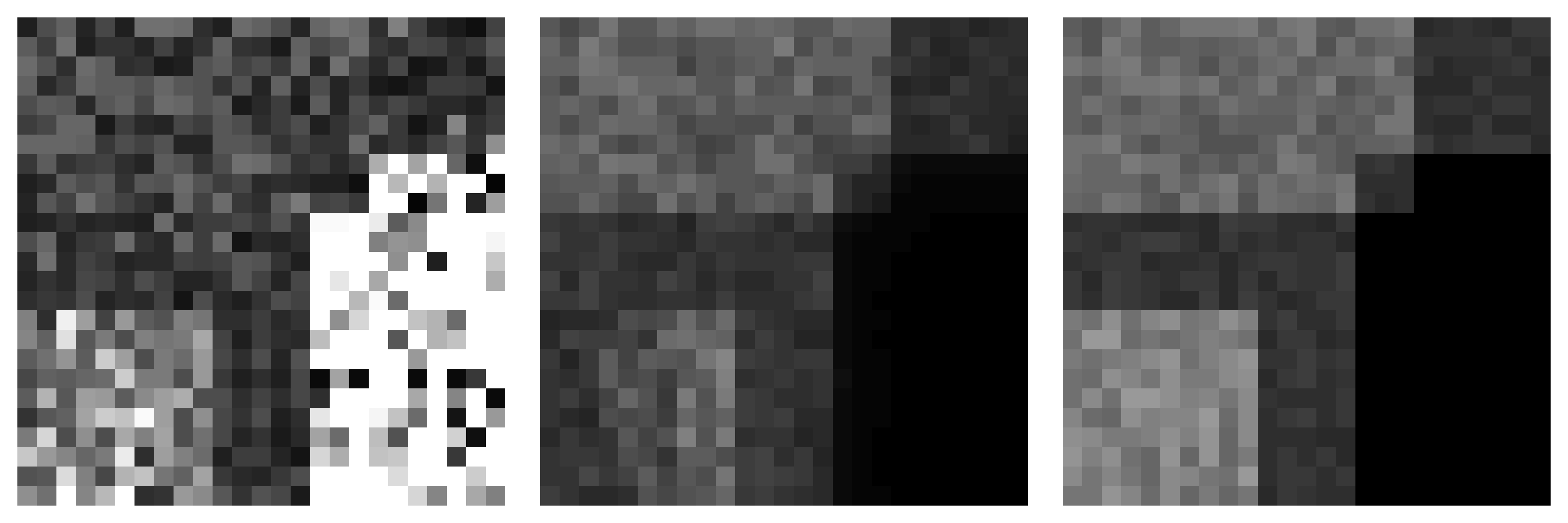}&
\includegraphics[scale=0.3]{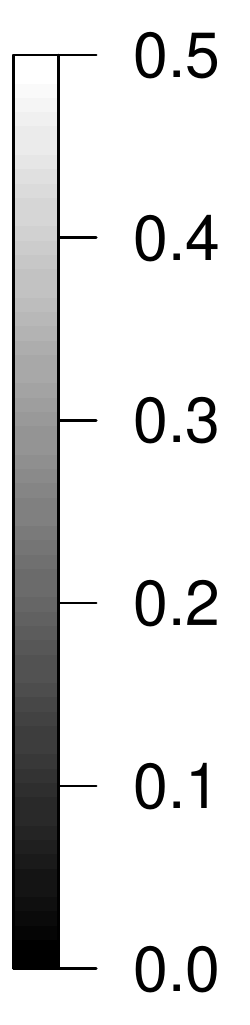}\\
\begin{rotate}{90}\hspace{10mm}$K^{\text{trans}}_{2}$\end{rotate}&
\includegraphics[scale=0.2]{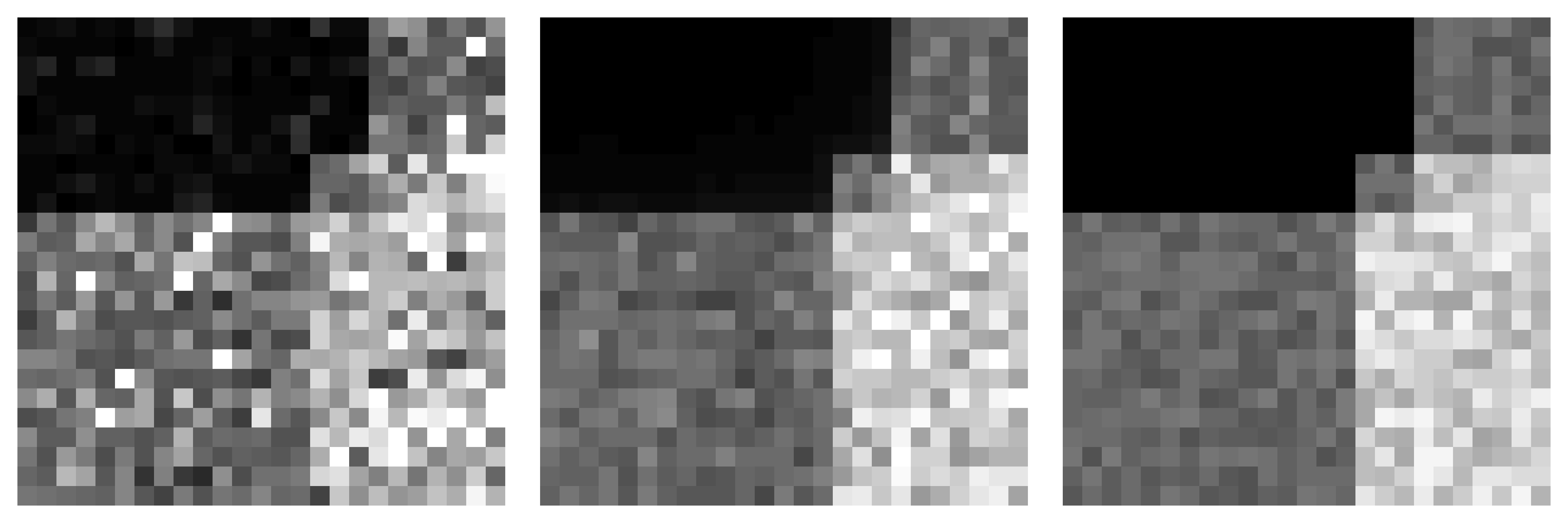}&
\includegraphics[scale=0.3]{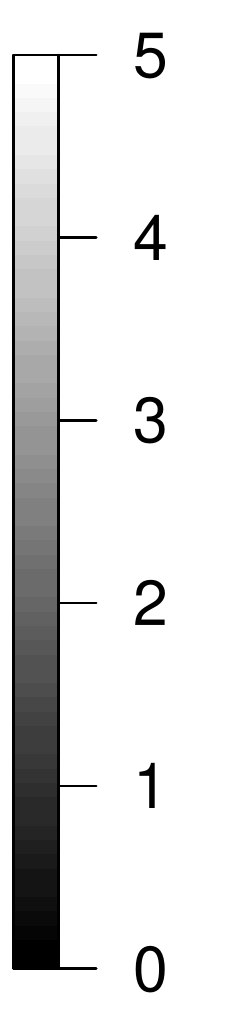}\\
\end{tabular}
\caption{Parameter maps for simulation study: Voxelwise (left column) and spatial fit (middle) of 2Comp model for true underlying values (right)}
\label{Fig:mapsSimulation}
\end{figure}

In Fig.~\ref{Fig:mapsSimulation} we show the parameter maps for the estimates of $k_{\text{ep}_1}$, $k_{\text{ep}_2}$, $K_1^{\text{trans}}$ and $K_2^{\text{trans}}$ 
from the voxelwise and the spatial model as well as the true underlying parameter values. As the voxelwise approach leads to unstable
estimates, the estimation results differ strongly from  the true underlying values. 
Especially for the voxels simulated from a 1Comp model, the voxelwise 2Comp model leads to unstable estimates.
For instance, for voxels in the lower right simulated from a true 1Comp model---with $v_{t_{1}}=0$---$k_{\text{ep}_1}$ is overestimated, $K_1^{\text{trans}}$ is overestimated, and consequently $K_2^{\text{trans}}$ is underestimated.
Compared to the voxelwise model, the spatial model leads to smooth parameter maps that can be interpreted more intuitively and to stable estimates that better match the true underlying parameter values.

In the spatial model, the parameter maps for $k_{\text{ep}_1}$ and $k_{\text{ep}_2}$ are smooth and the estimates match the true underlying values quite well.
There is some oversmoothing such that the higher $k_{\text{ep}_1}$ values in the  lower left corner are underestimated and as a consequence also the corresponding $k_{\text{ep}_2}$ are underestimated.
The estimates of $K_1^{\text{trans}}$ and $K_2^{\text{trans}}$ perfectly match the true underlying values.
For the blocks simulated from a 1Comp model either the $K_1^{\text{trans}}$ estimate or the $K_2^{\text{trans}}$ estimate becomes zero.
Like this, model redundancy is avoided and the posteriors are unimodal.

\section{Breast cancer study}

\subsection{Data description}
To evaluate the clinical use of our approach we use a previously 
analysed DCE-MRI study on breast cancer~\citep{schmid06}. The dataset 
consists of twelve patients with breast tumour. The scans were acquired once before and once after 
six cycles of chemotherapy. The drug is expected to stop the process of angiogenesis, \ie, to lower 
the elevated blood flow to the tumour, hence, to lower the kinetic parameters $K^{\text{trans}}$ and $k_{\text{ep}}$. As clinical evaluation, tumours were removed at the end of 
therapy and the response to therapy was evaluated histologically. Six of the twelve patients were identified as responders, the other six as nonresponders. Informed consent was obtained from all patients and the study was acquired in accordance with recommendations given by \cite{Leach2005}.

The scans were acquired with a 1.5 T Siemens MAGNETOM
Symphony scanner, \textit{TR} $=11$ $ms$  and \textit{TE} $=4.7$ $ms$. Each scan consists of
three slices of $230 \times 256$ voxels, but only the central slice was used in our analysis. 
A dose of $D = 0.1$ $mmol/kg$ body weight Gd-DTPA was injected at the start of the fifth acquisition using a power injector. 
Regions of interest cover the tumour and surrounding normal tissue.

\subsection{Results}

\begin{figure}[t]
\begin{tabular}{cccccc}
& \multicolumn{2}{c}{\textbf{patient 4}} & \multicolumn{2}{c}{\textbf{patient 6}}\\
& pre & post & pre & post & \\ 
\begin{rotate}{90}\hspace{10mm}$k_{\text{ep}_1}$\end{rotate} &
\includegraphics[scale=0.15]{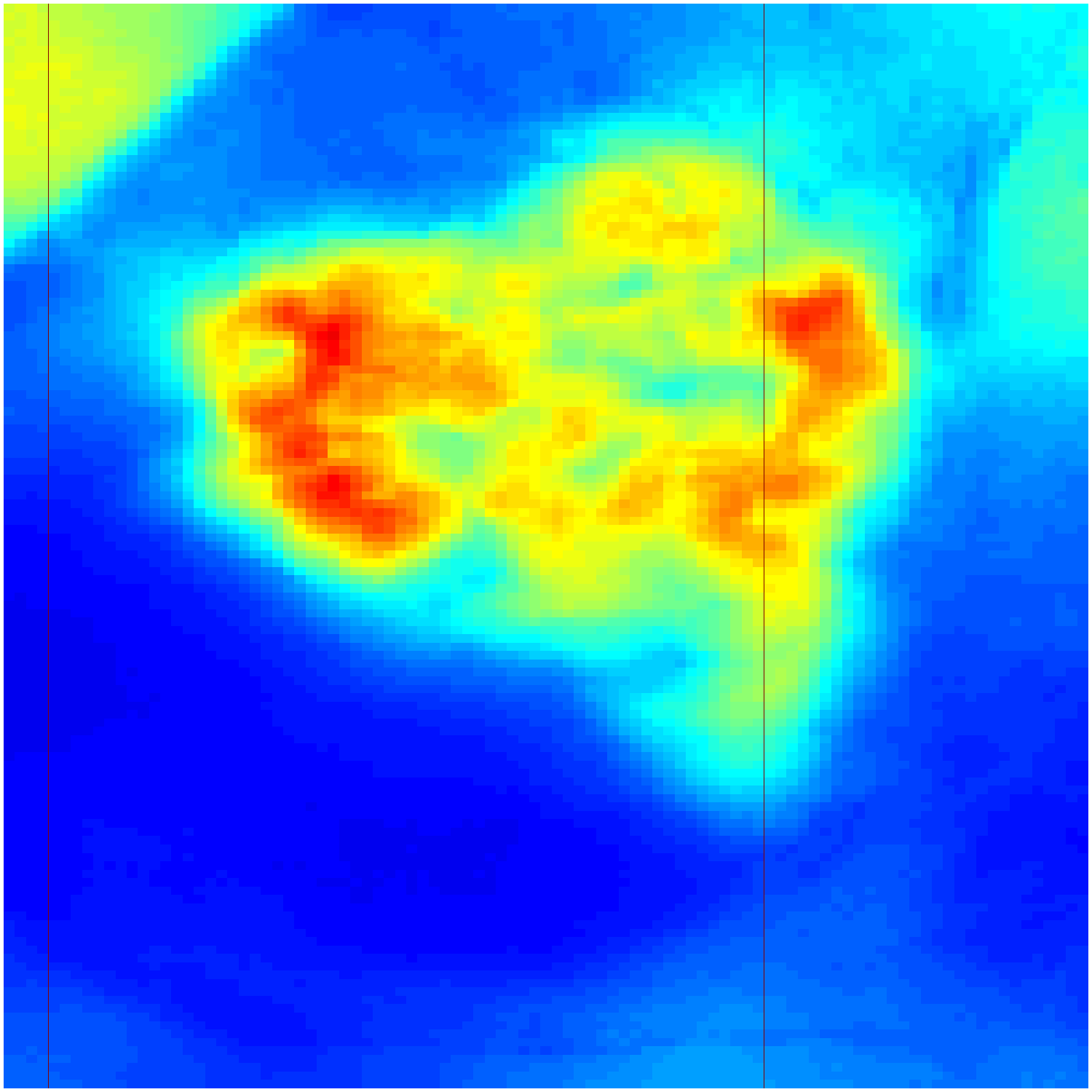} &
\includegraphics[scale=0.15]{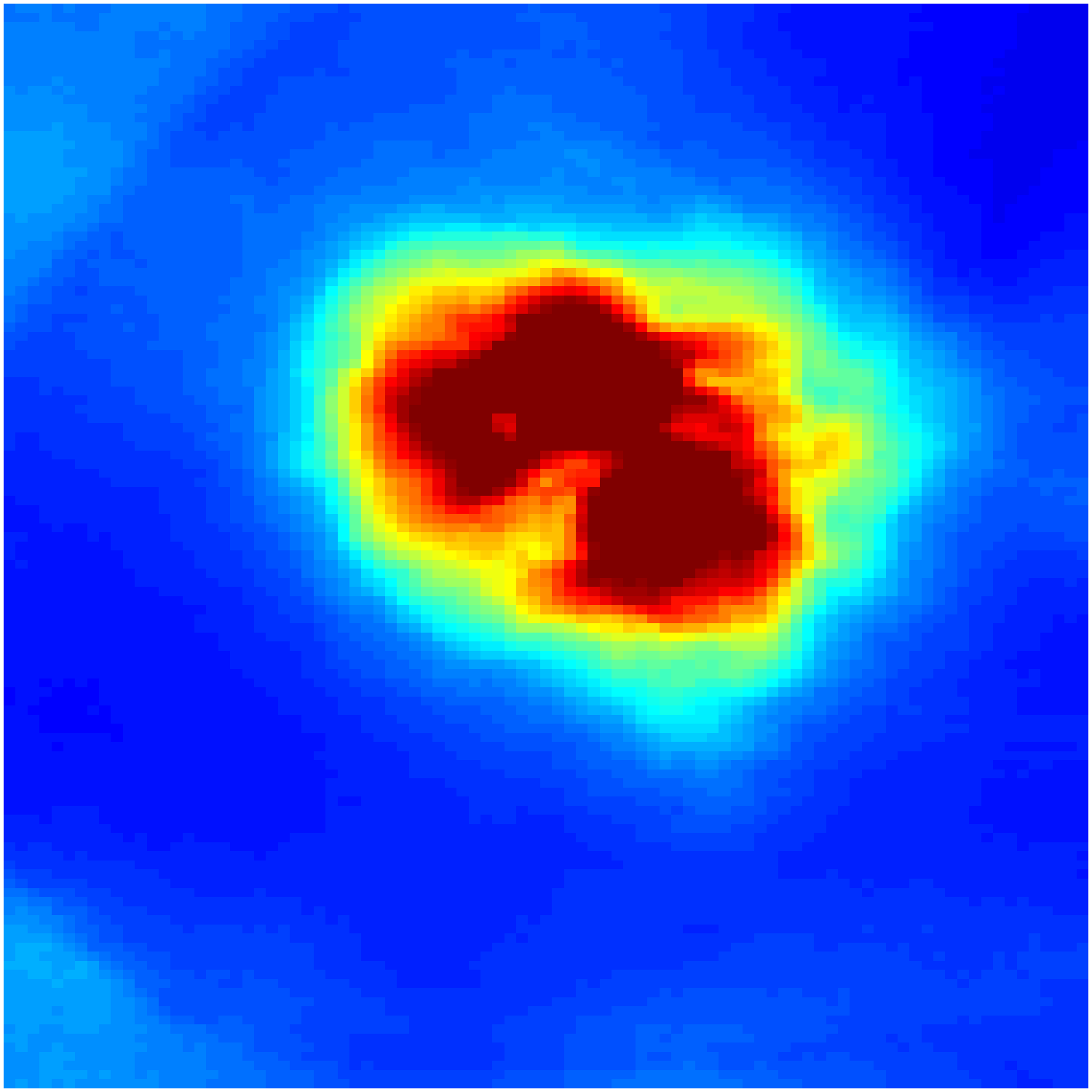} &
\includegraphics[scale=0.15]{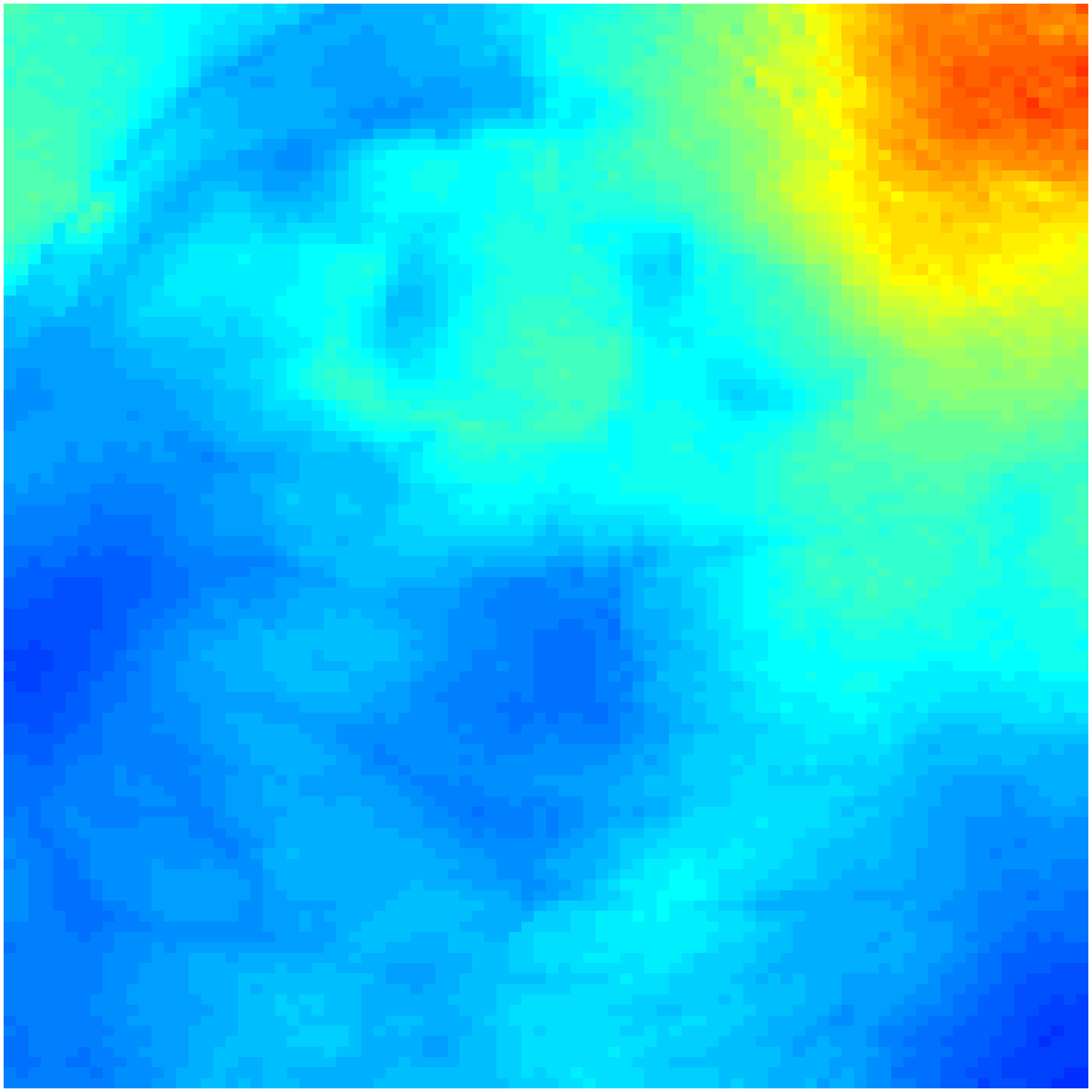} &
\includegraphics[scale=0.15]{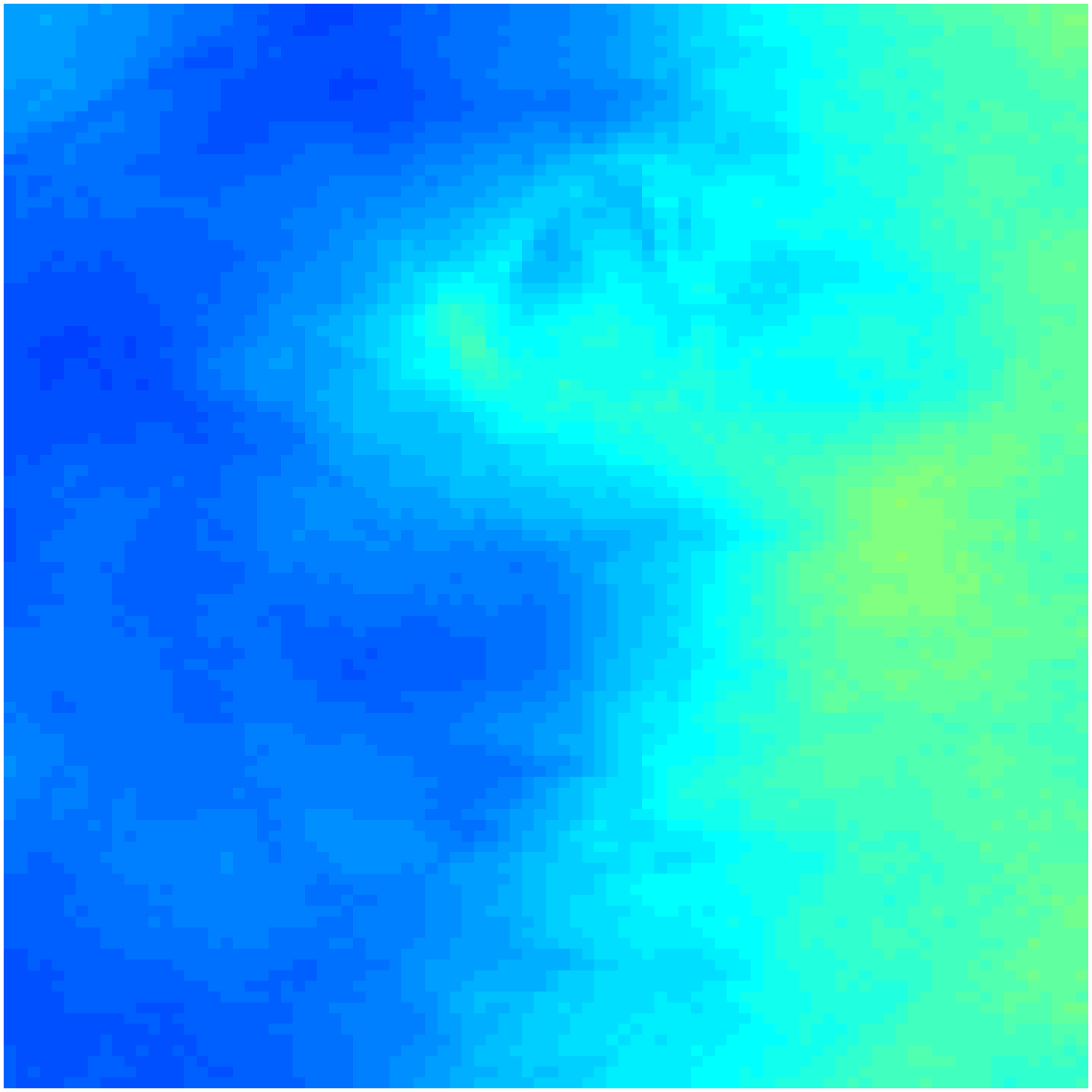} &
\includegraphics[scale=0.25]{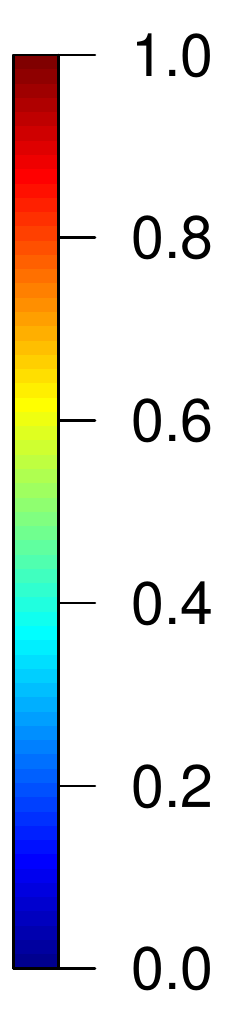}\\
\begin{rotate}{90}\hspace{10mm}$k_{\text{ep}_2}$ \end{rotate}&
\includegraphics[scale=0.15]{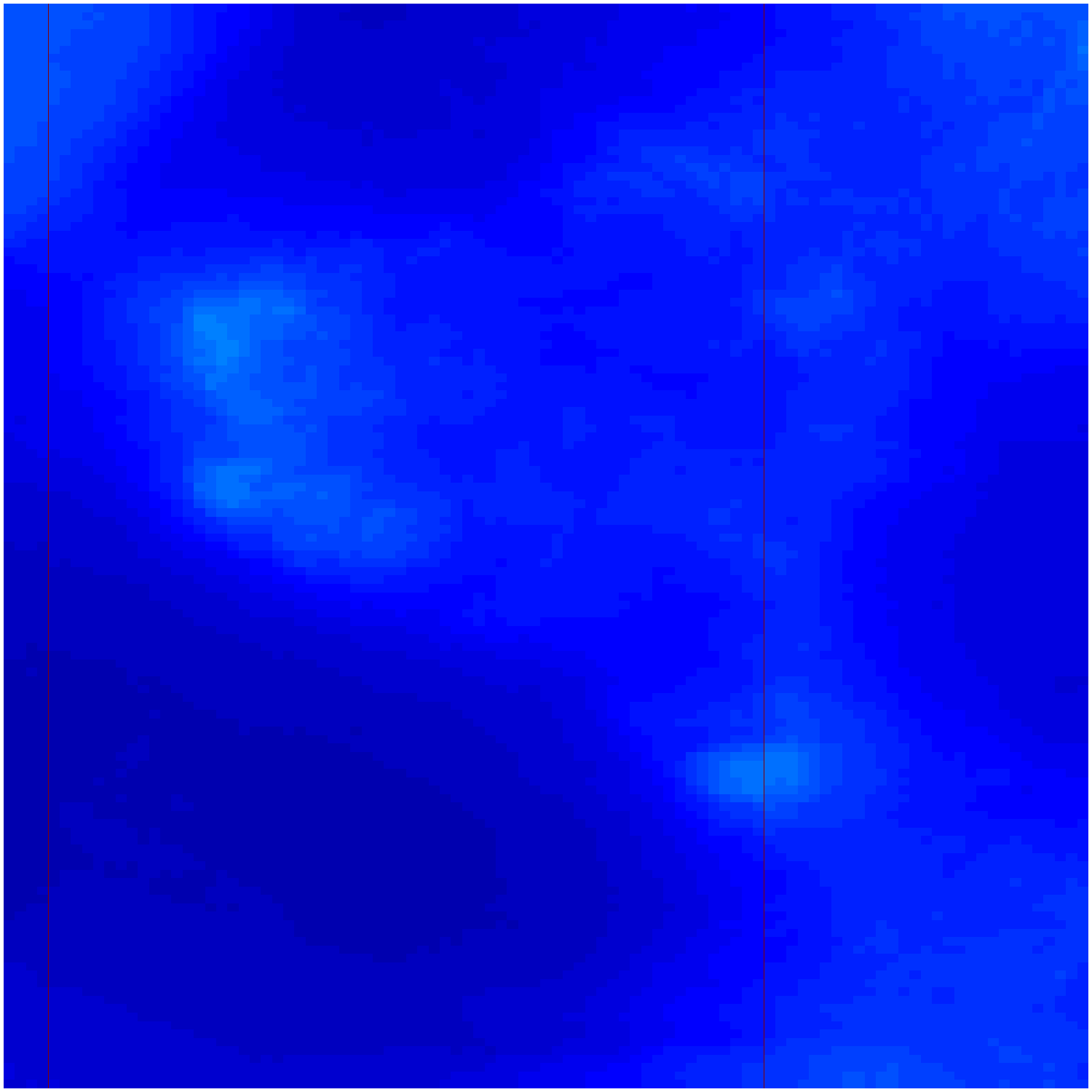} &
\includegraphics[scale=0.15]{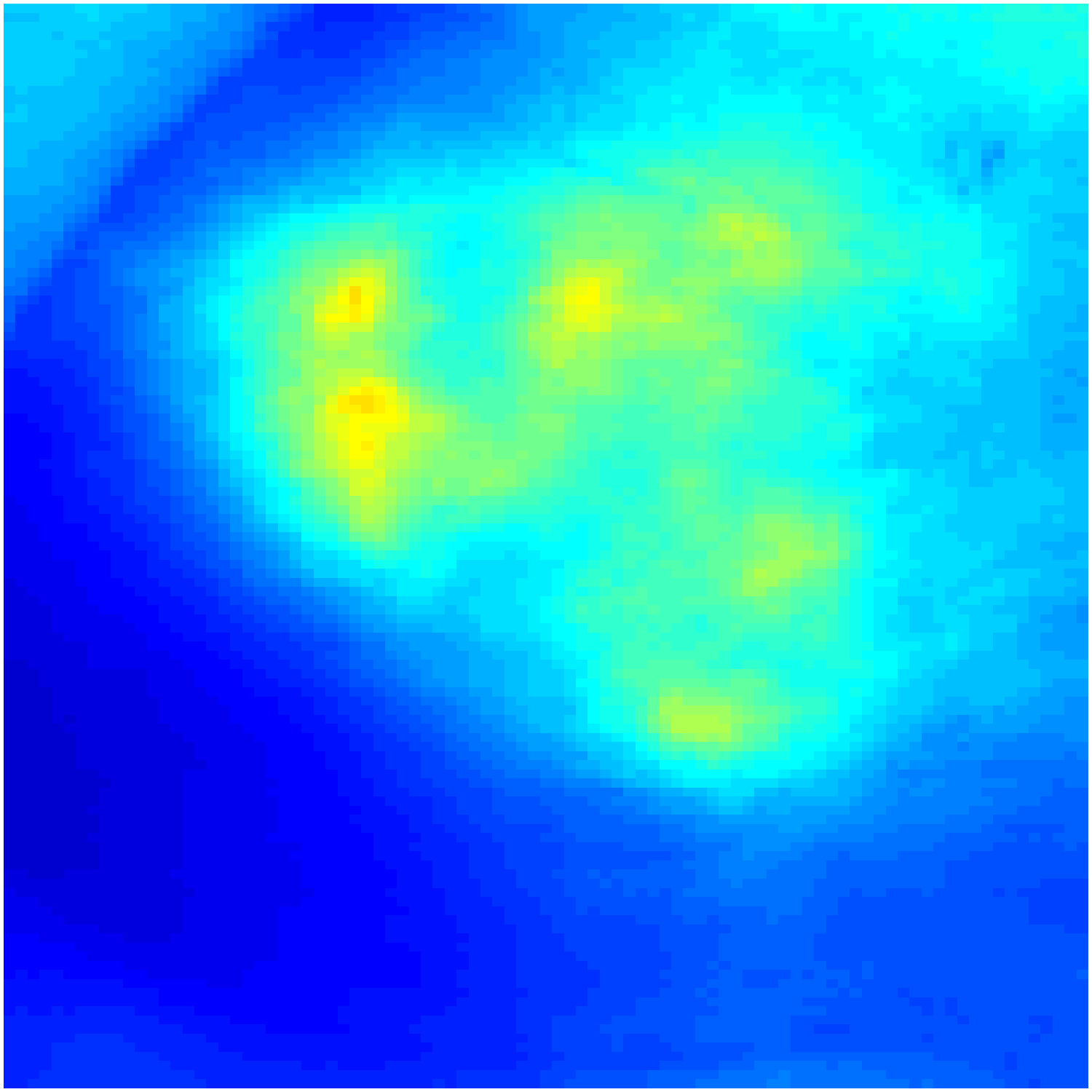} &
\includegraphics[scale=0.15]{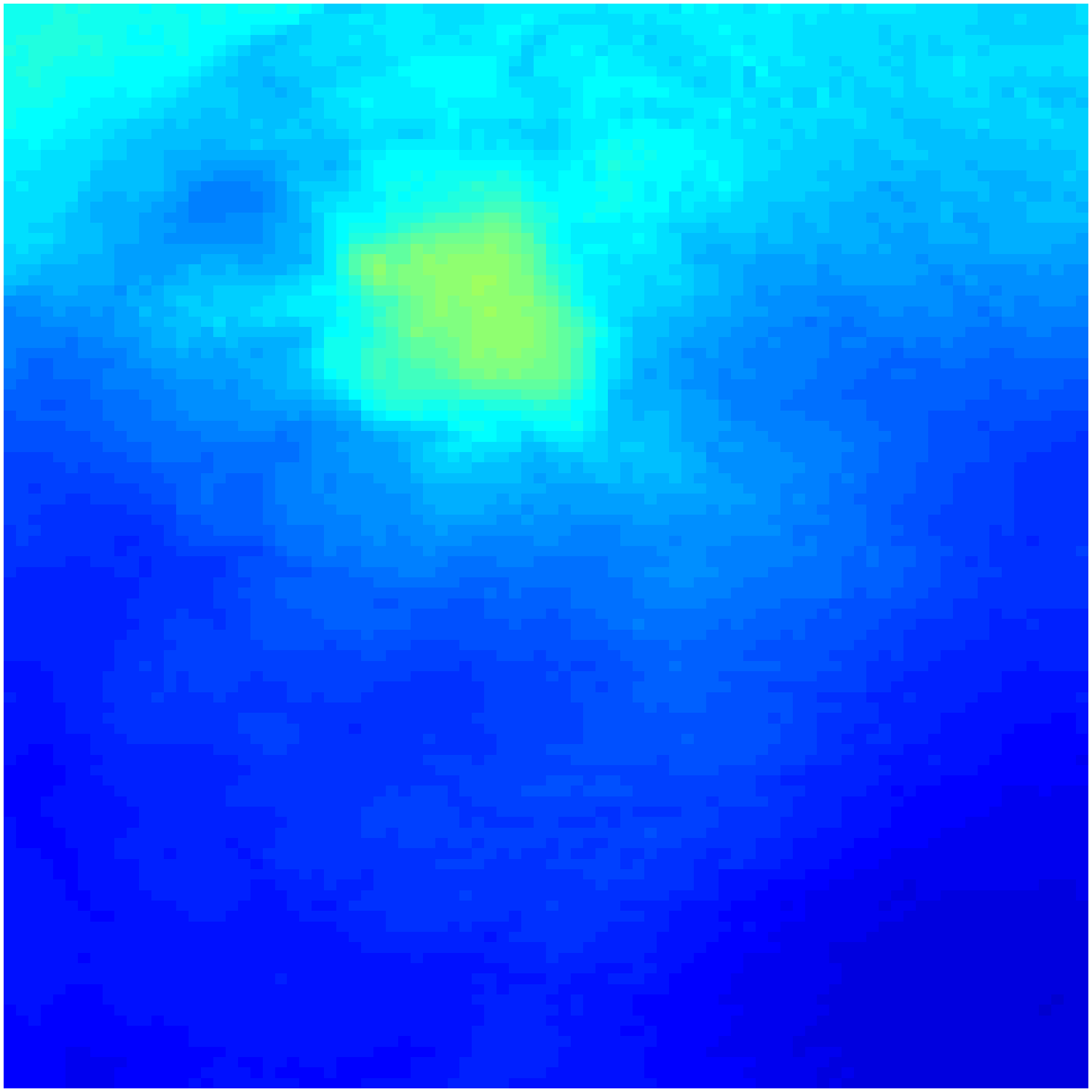} &
\includegraphics[scale=0.15]{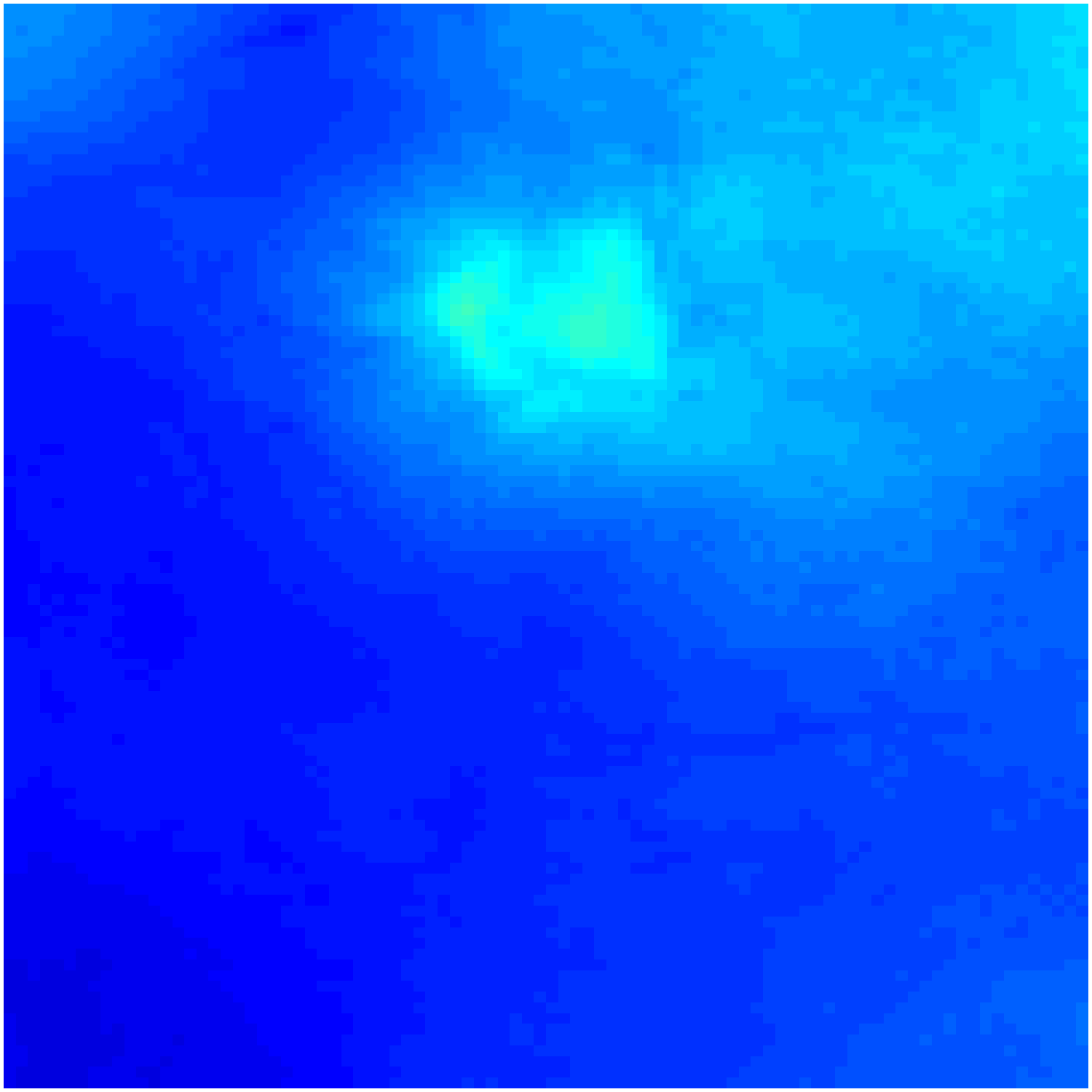} &
\includegraphics[scale=0.25]{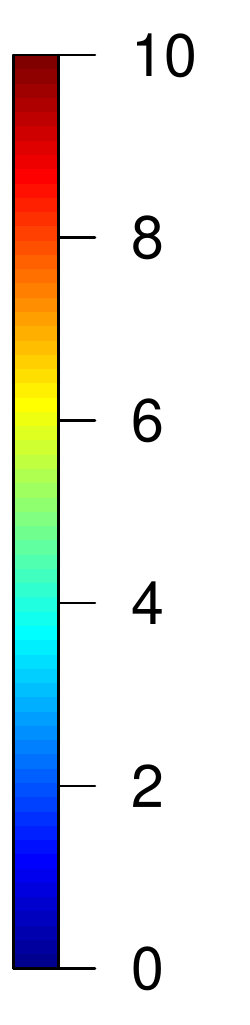}\\
\begin{rotate}{90}\hspace{10mm}$K^{\text{trans}}_{1}$ \end{rotate}&
\includegraphics[scale=0.15]{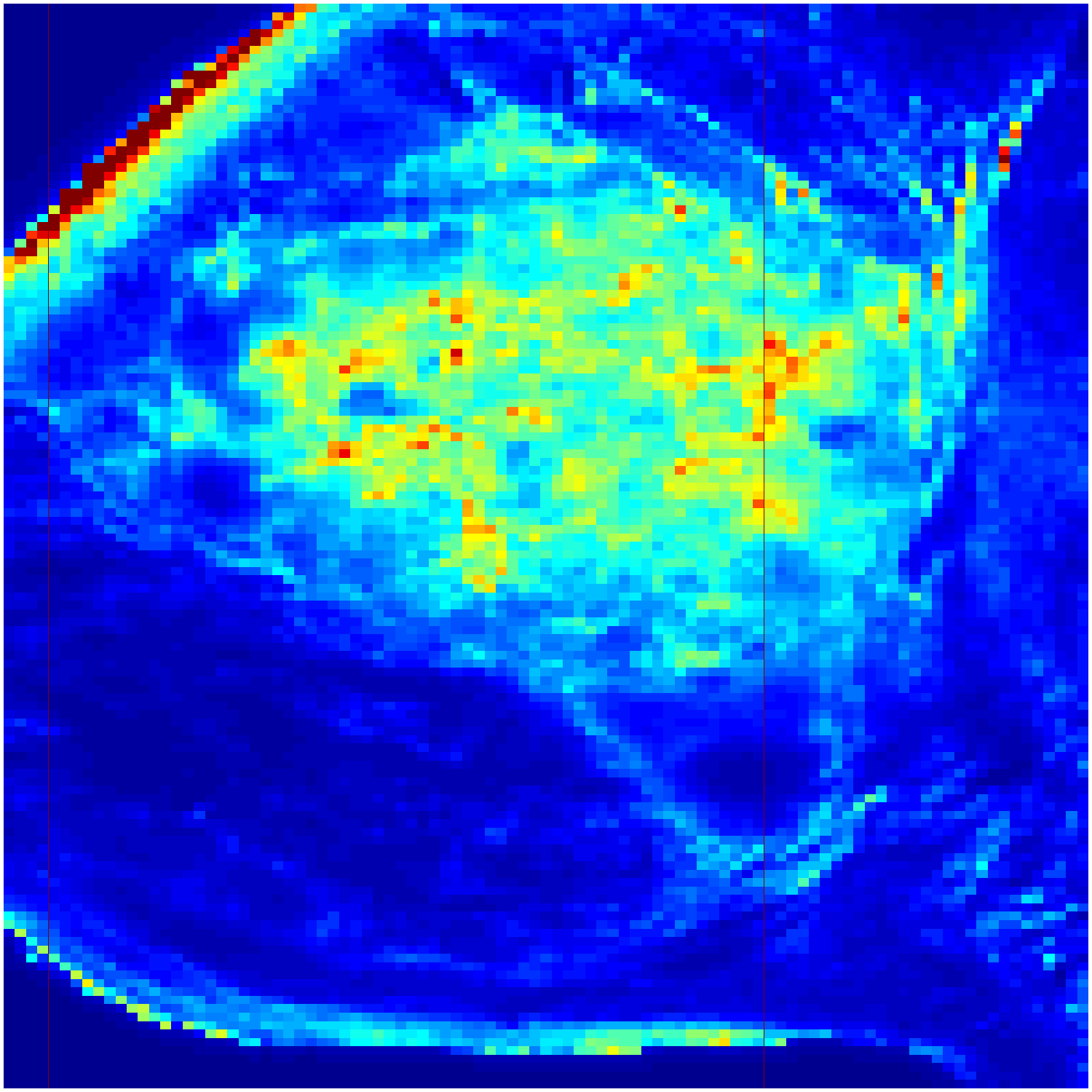} &
\includegraphics[scale=0.15]{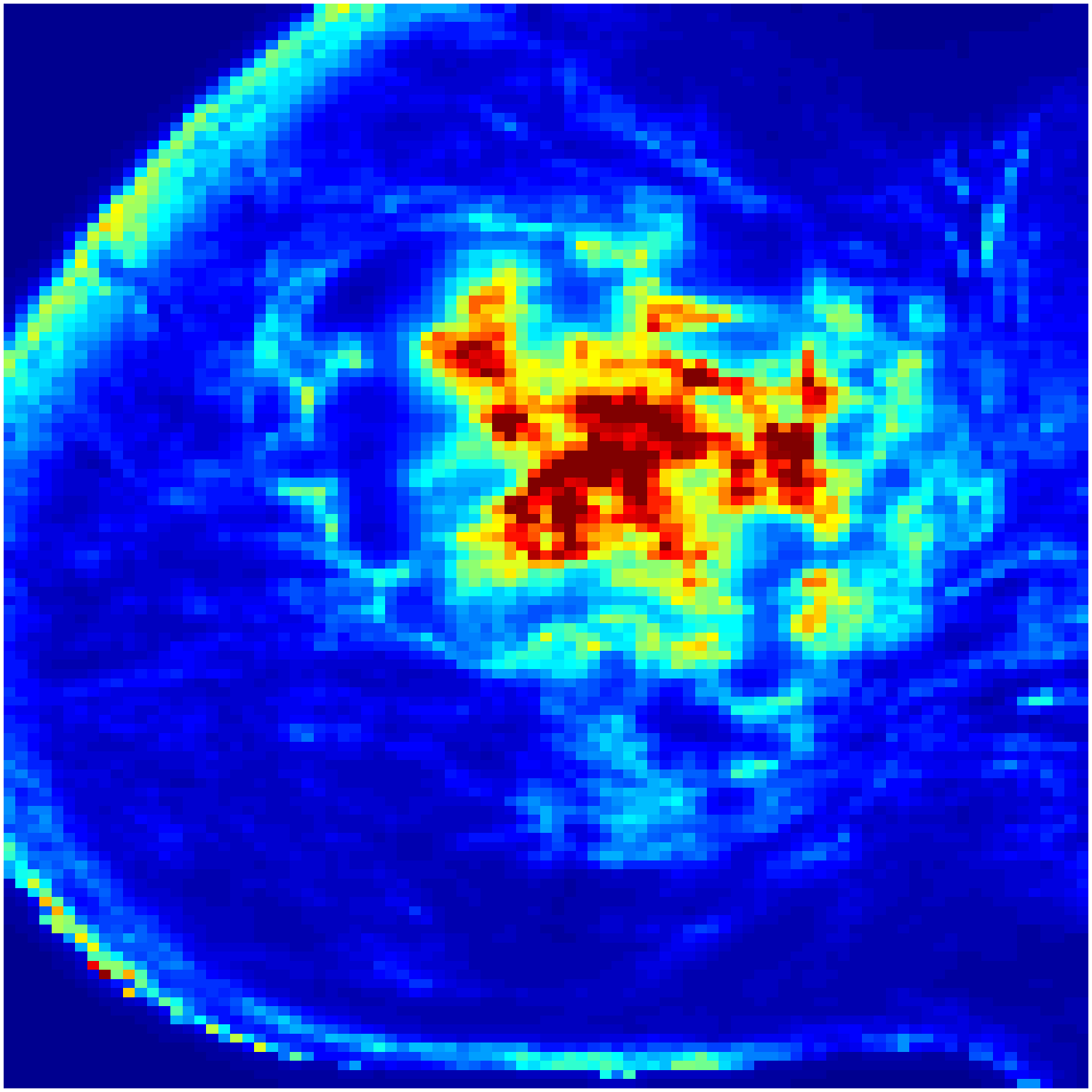} &
\includegraphics[scale=0.15]{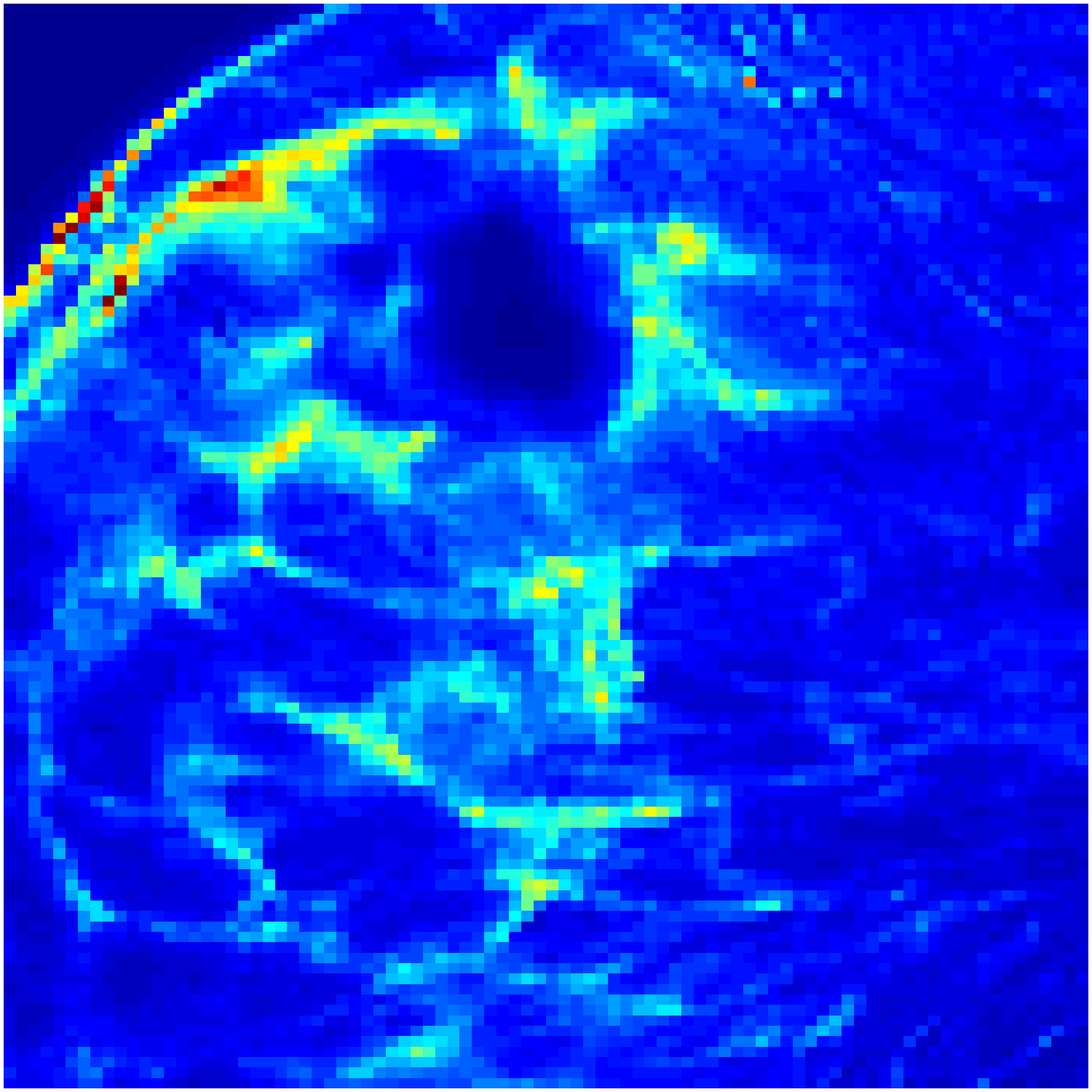} &
\includegraphics[scale=0.15]{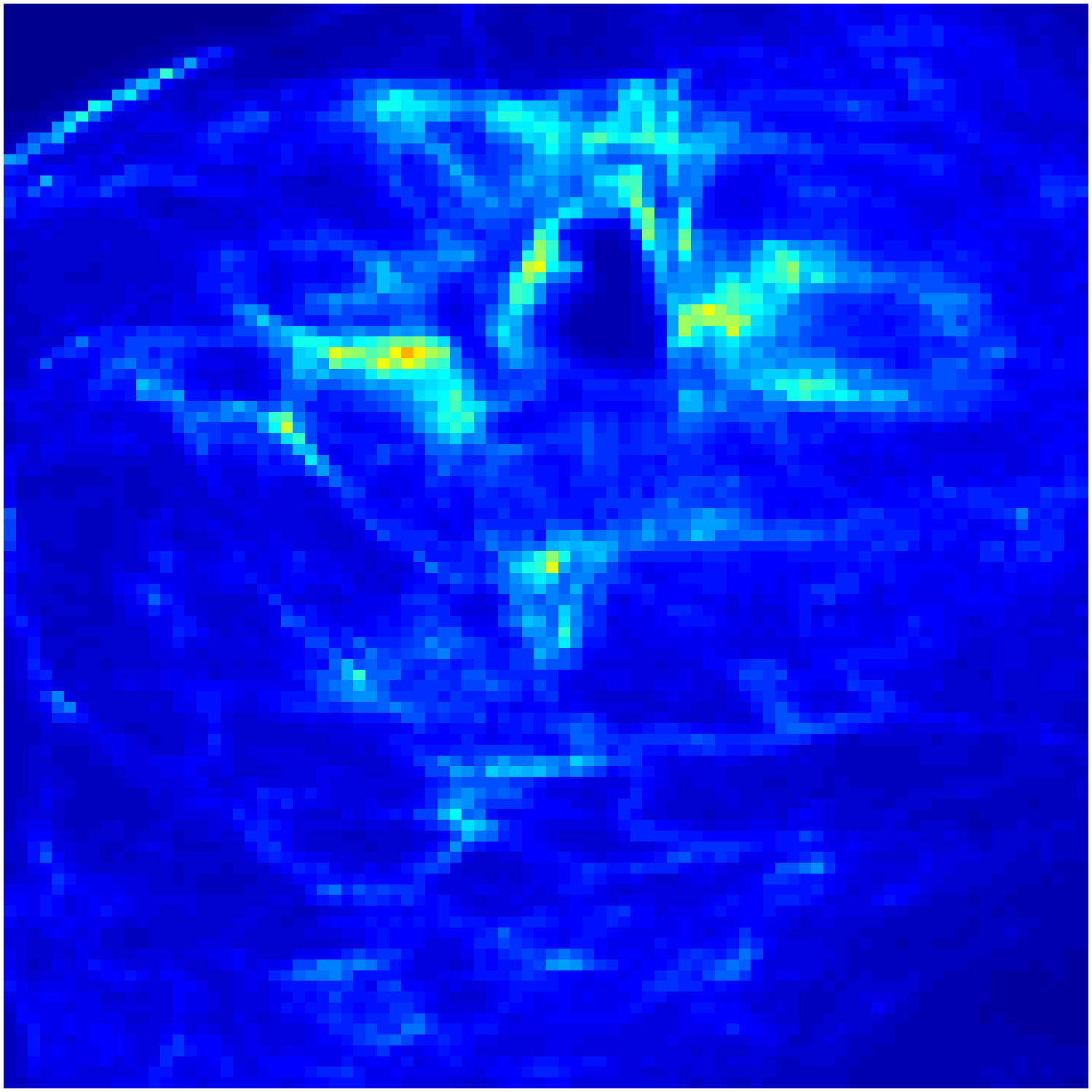} &
\includegraphics[scale=0.25]{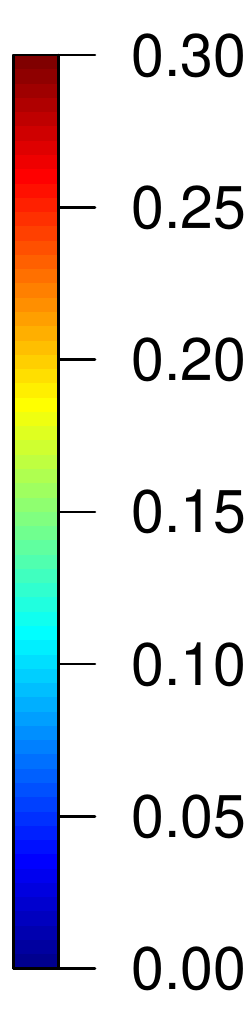}\\
\begin{rotate}{90}\hspace{10mm}$K^{\text{trans}}_{2}$ \end{rotate}&
\includegraphics[scale=0.15]{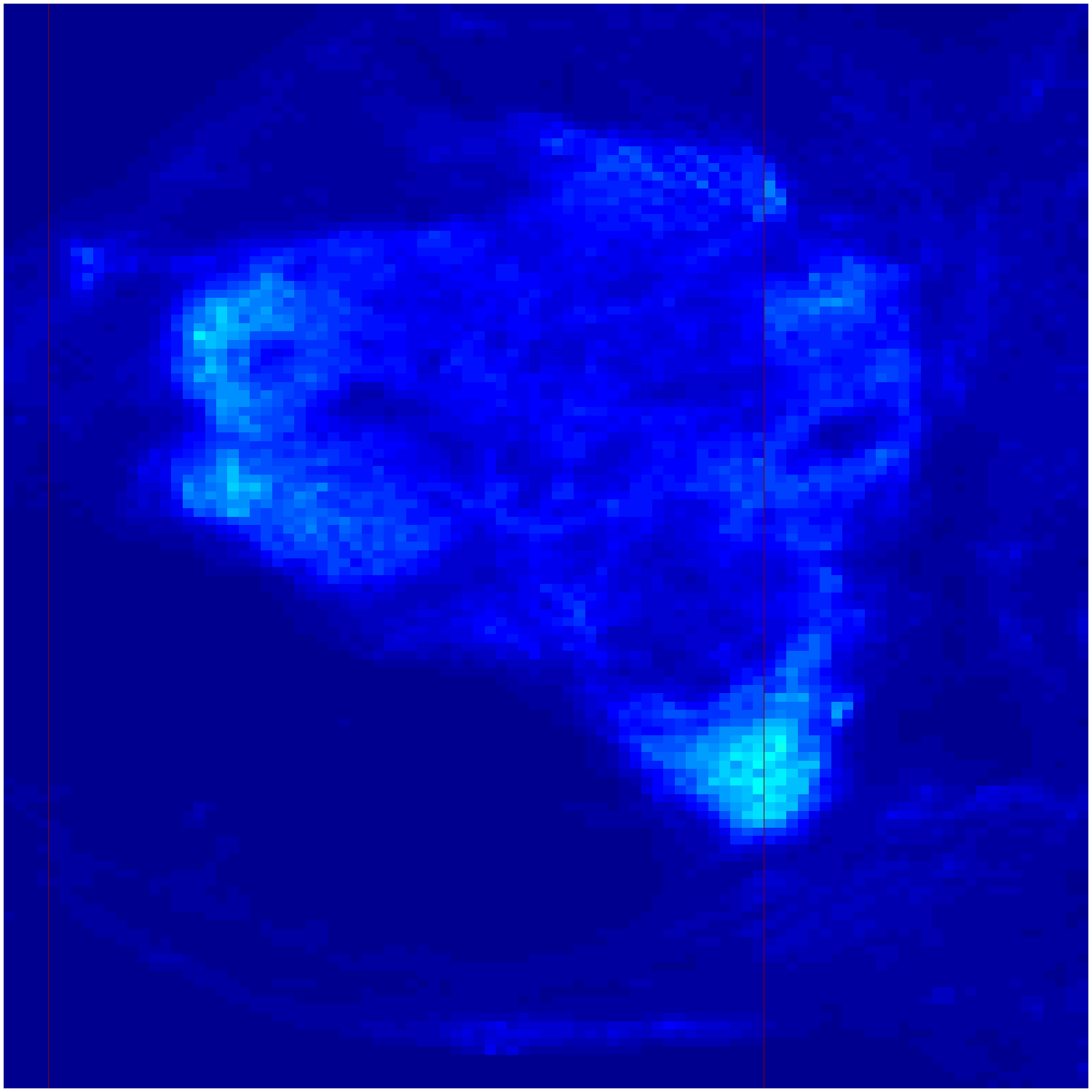} &
\includegraphics[scale=0.15]{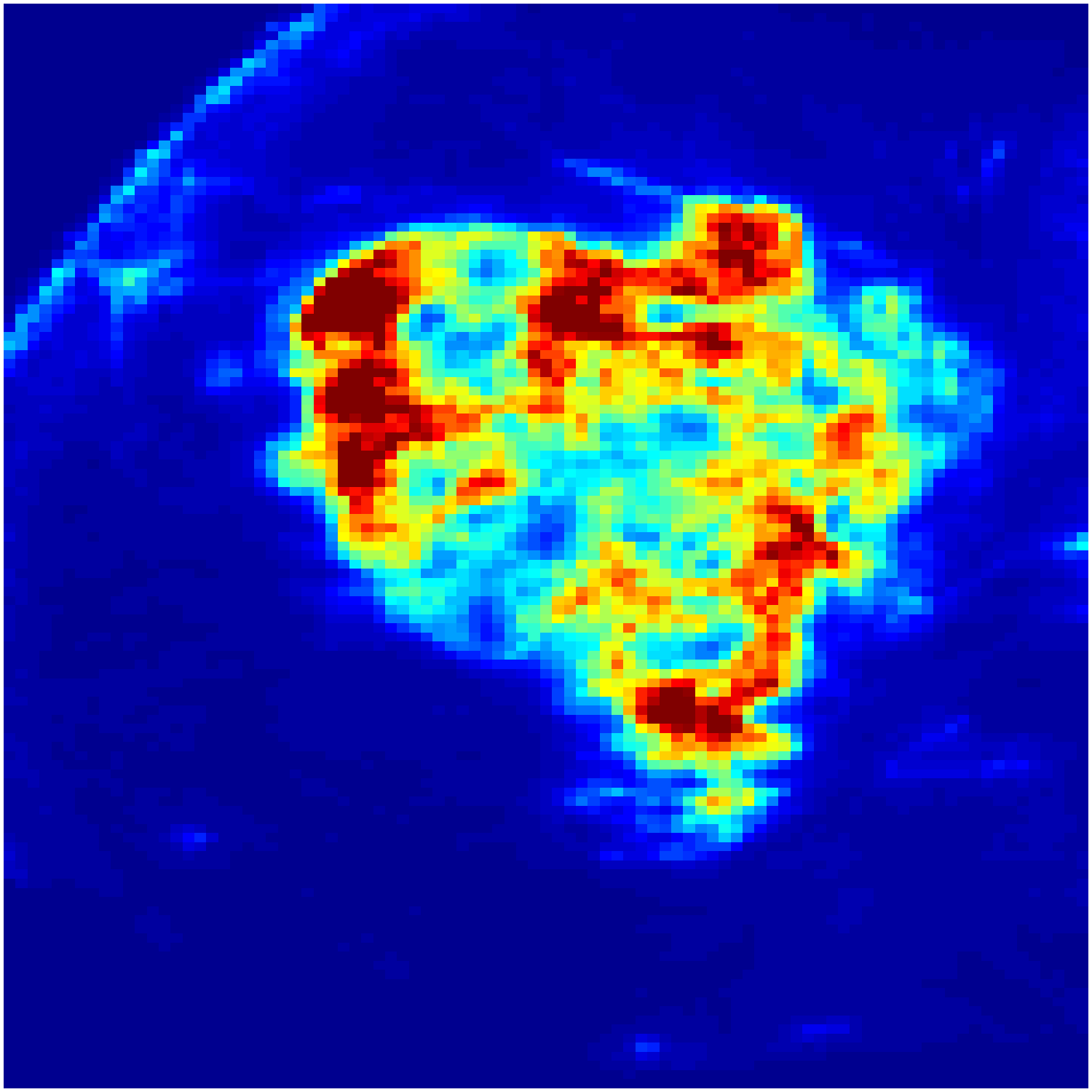} &
\includegraphics[scale=0.15]{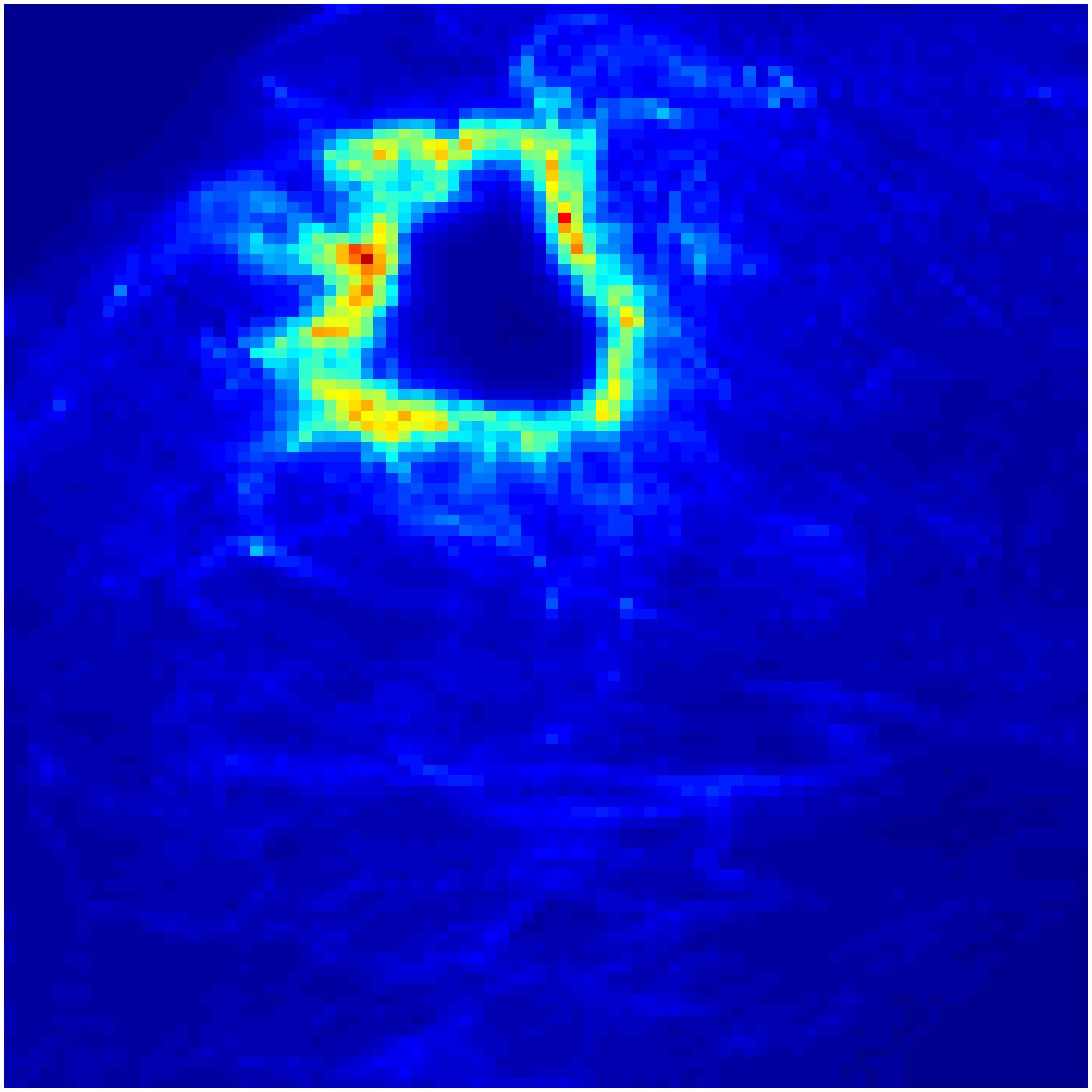} &
\includegraphics[scale=0.15]{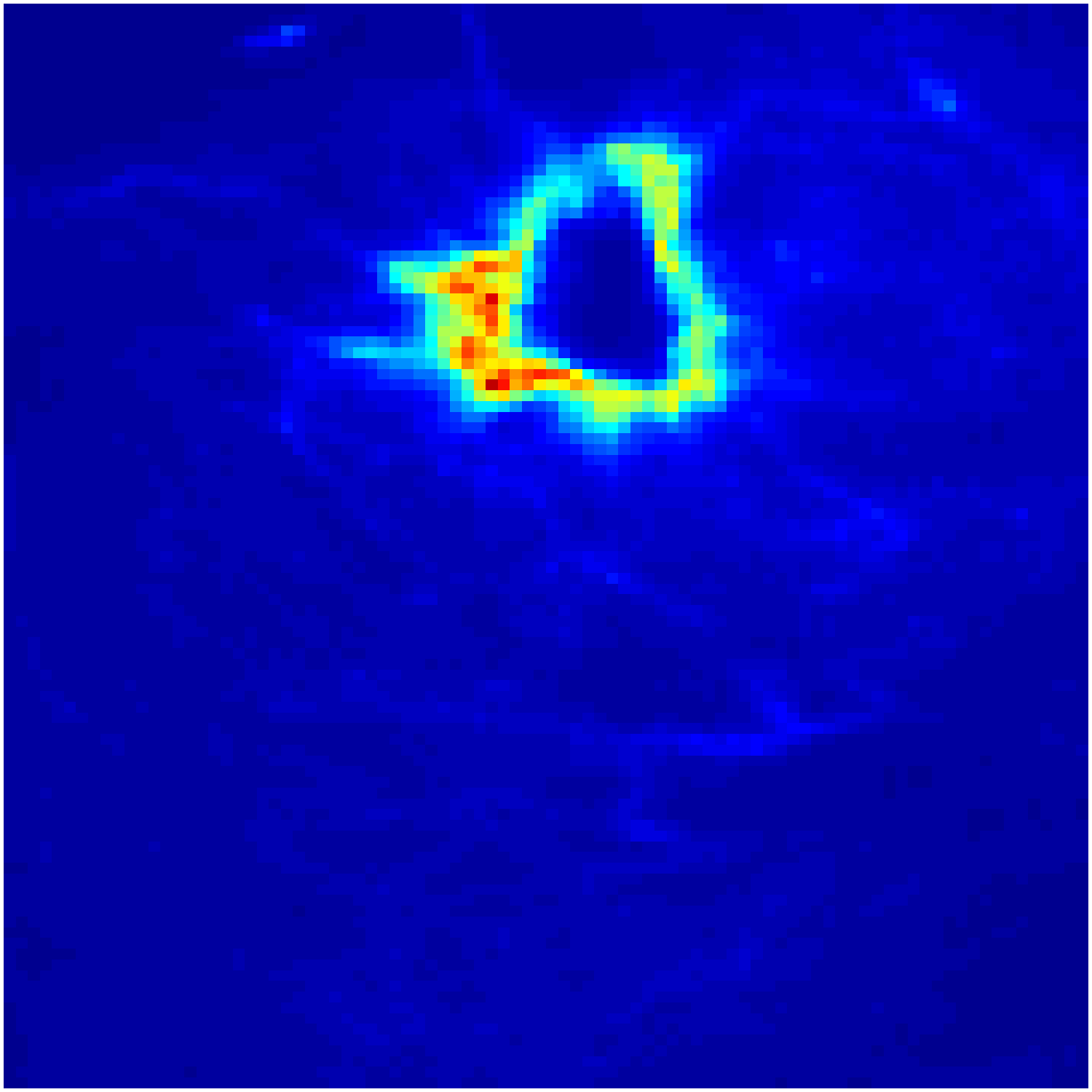} &
\includegraphics[scale=0.25]{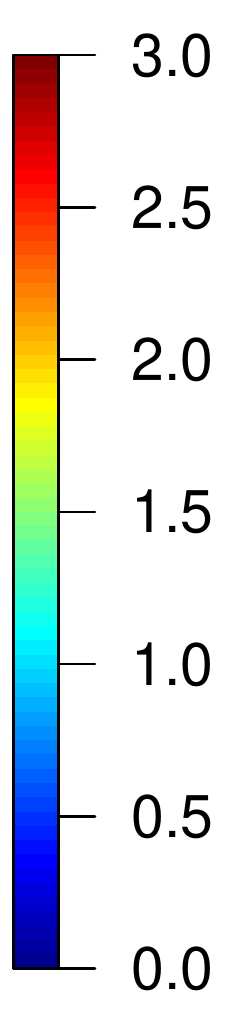}\\
\begin{rotate}{90}\hspace{10mm}$p_D$\end{rotate} &
\includegraphics[scale=0.15]{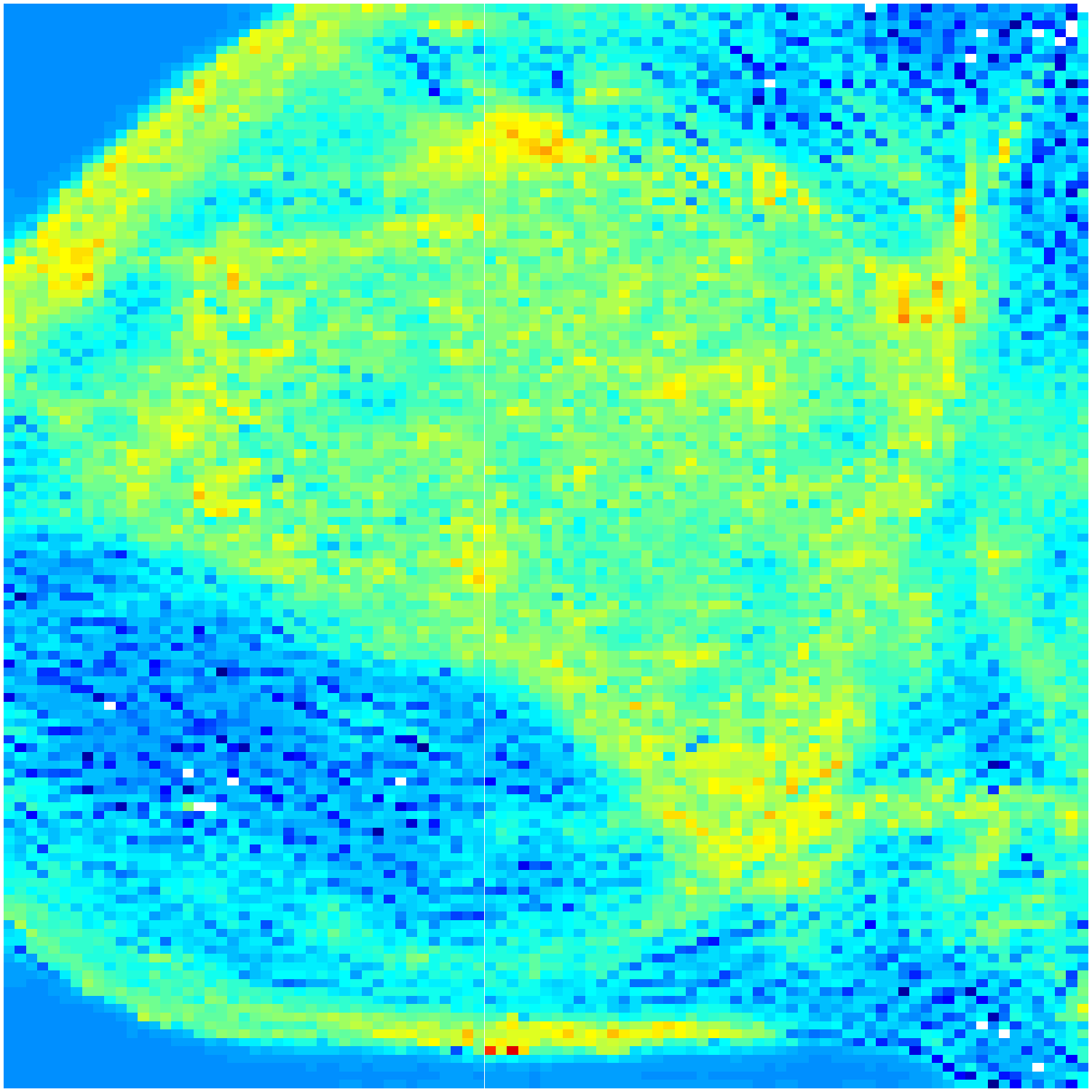}&
\includegraphics[scale=0.15]{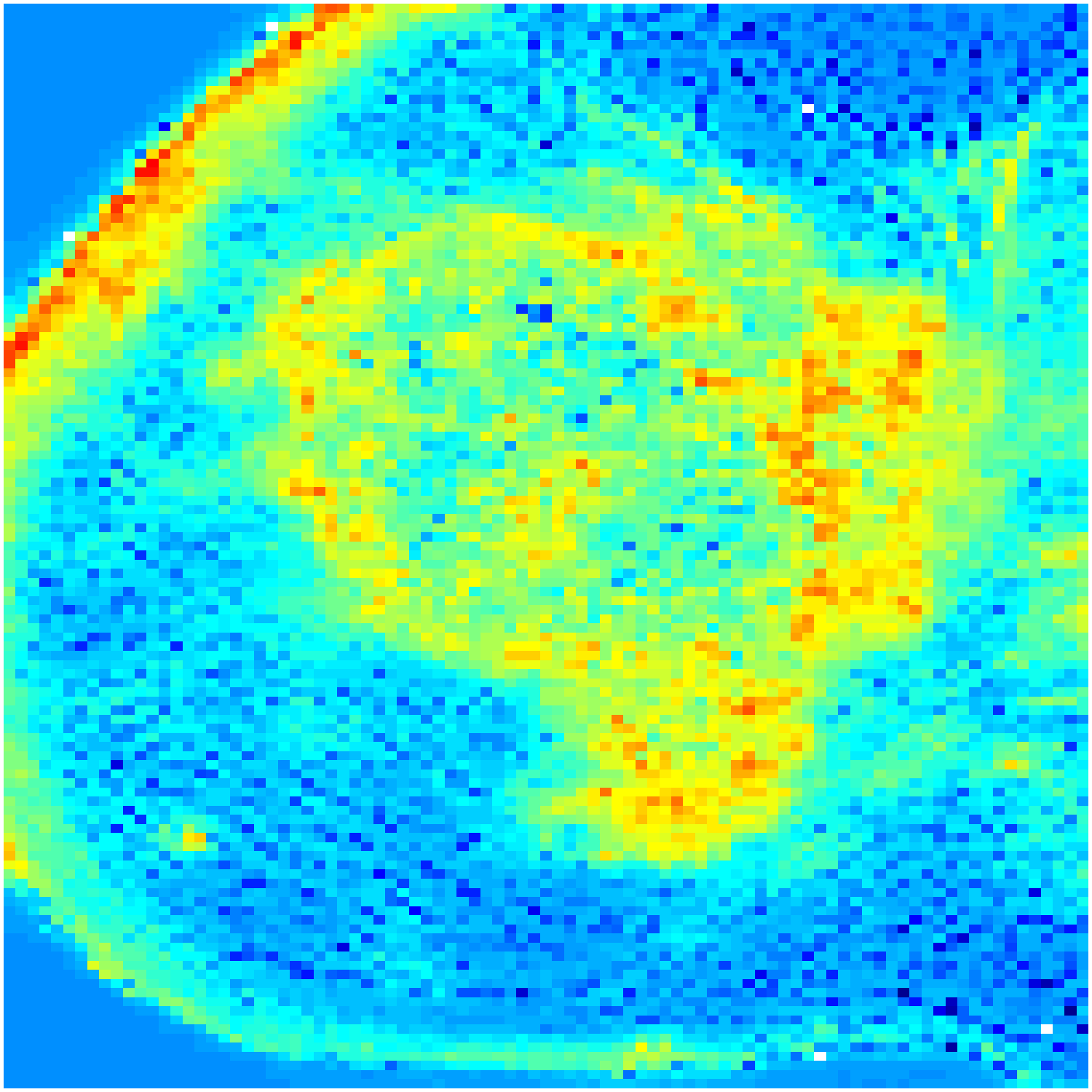}&
\includegraphics[scale=0.15]{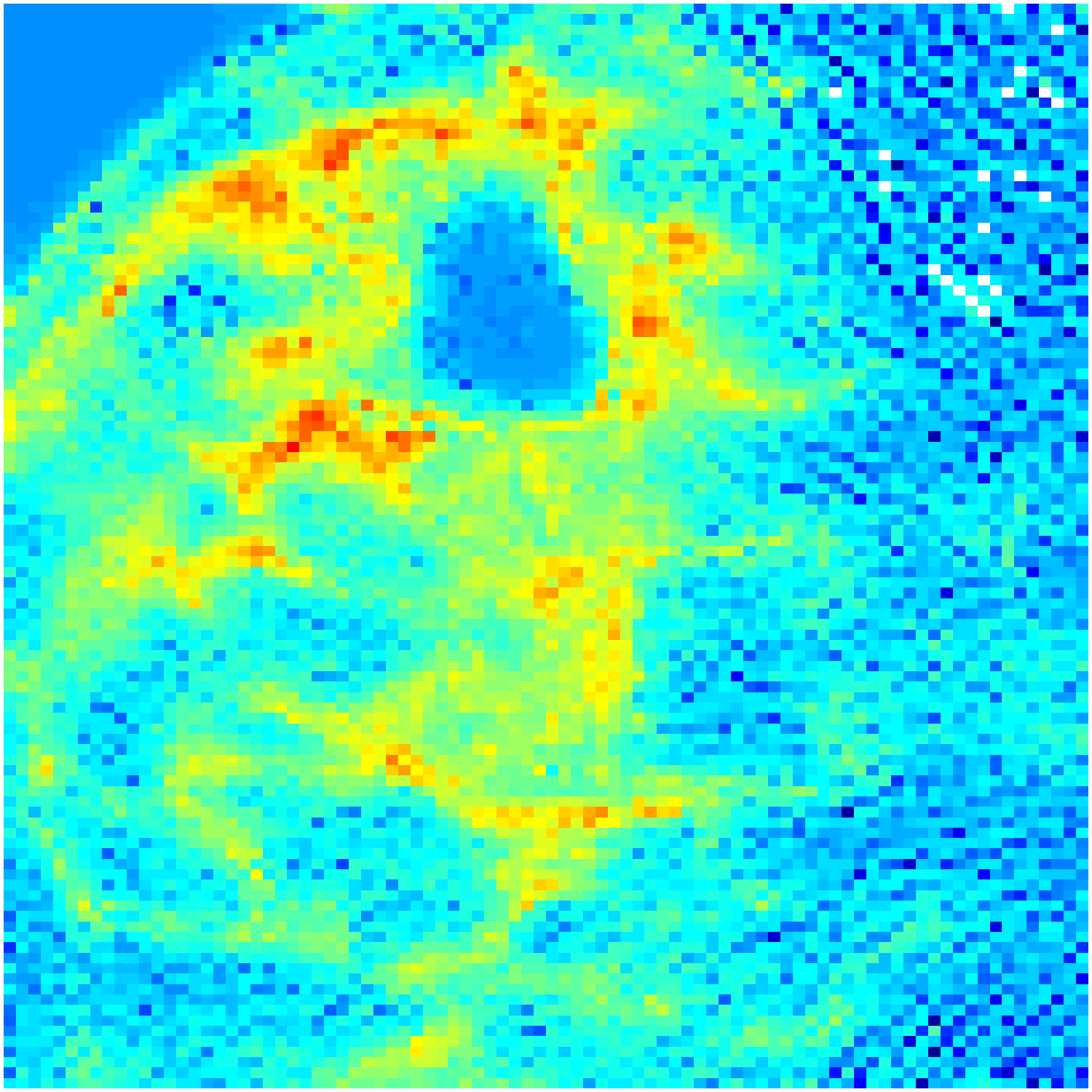}&
\includegraphics[scale=0.15]{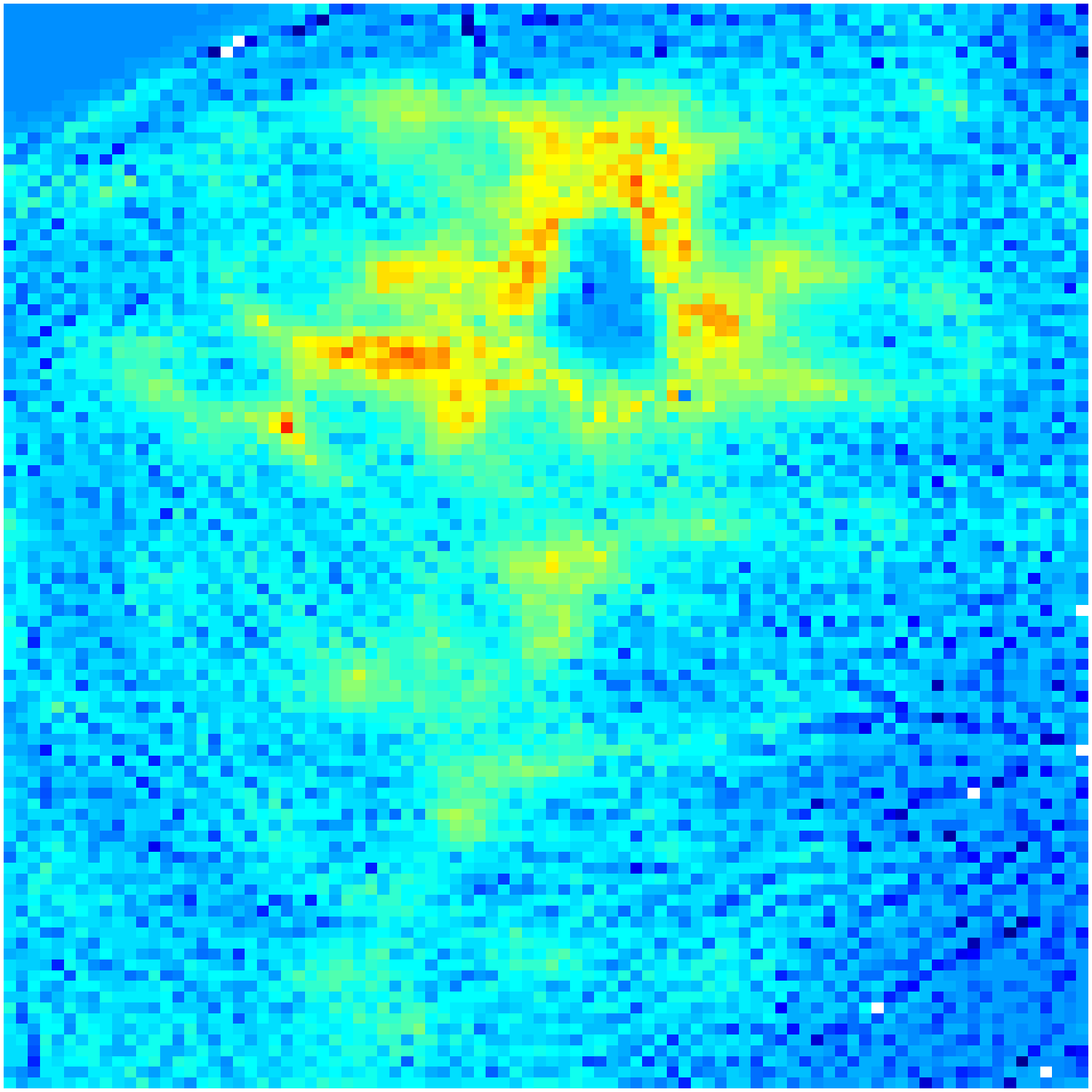}&
\includegraphics[scale=0.25]{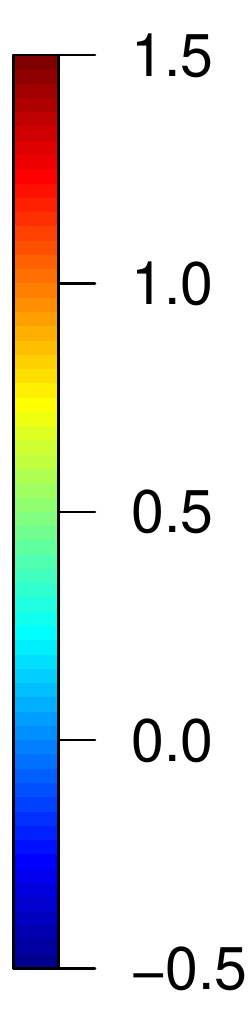}
\end{tabular}
\caption{Spatial 2Comp model: parameter maps for patients 4 and 6 pre- and post-treatment}
\label{Fig:MapsBreast}
\end{figure}

In \fref{Fig:MapsBreast} the parameter maps from the spatial 2Comp model are shown for pre- and post-treatment scans of patient 4 (nonresponder to therapy) and patient 6 (responder).
Similar to the results of the simulation study and in accordance with the prior assumptions, the estimated parameter maps for the DCE-MR images are quite smooth for the exponential rates $k_{\text{ep}_1}$ and $k_{\text{ep}_2}$, whereas the $K_1^{\text{trans}}$ and $K_2^{\text{trans}}$ estimates show more spatial variation. 
The contribution of the second compartment vanishes ($K_2^{\text{trans}}$ close to zero) in healthy tissue. In those regions, the 1Comp model suffices to describe the observed uptake dynamics, meaning that the tissue is homogeneous there. Interestingly, tissue inside of the tumour is often homogeneous as well.
The second compartment has nonzero contribution and improves the fit of observed CTCs at tumour margins and in parts of the surrounding tissue.
In those regions, the tissue is heterogeneous as both the slow and the fast exchanging compartments contribute to the uptake dynamics.  Larger $p_D$ values and improved fit compared to the 1Comp model reflect this heterogeneity.

For patient 4 the parameter maps for the pre-treatment scan depict high $k_{\text{ep}_1}$ and $k_{\text{ep}_2}$ values as well as high $K_1^{\text{trans}}$ and $K_2^{\text{trans}}$ values for a large tissue region. Post-treatment, the kinetic parameters $k_{\text{ep}_1}$, $k_{\text{ep}_2}$, $K_1^{\text{trans}}$ and $K_2^{\text{trans}}$ have higher values, but the tissue region with increased blood flow becomes smaller and more dense.
Reduced tumour volume could easily be misinterpreted as treatment success.
Here, the $p_D$ map contains additional information that might help to assess treatment success. For patient 4 voxelwise $p_D$ values are even higher in the post-treatment scan.

For patient 6 parameter maps  of $k_{\text{ep}_1}$ and $k_{\text{ep}_2}$ are quite smooth. The contribution of the second compartment---$K_2^{\text{trans}}$---is close to zero inside of the tumour and in surrounding healthy tissue, and it is not vanishing at tumour margins and in surrounding tissue. After treatment, the number of voxels where the second compartment contributes decreases notably.
For patient 6 tumour margins and extensions around the tumour are heterogeneous and better described with the aid of an additional second compartment. 
Both the $p_D$ values and the size of the tissue region with increased $p_D$ decrease after treatment.

\section{Conclusions}
In this paper we have discussed redundancy issues of a specific nonlinear regression problem, namely the estimation of kinetic parameters in a two tissue compartment model for DCE-MR images.
With a spatial prior we have regularised the parameter space and made the parameters identifiable.
With this prior, a 2Comp model can be fitted at a voxel level and CTCs in heterogeneous tissue, especially at tumour margins, can be described better than with the standard 1Comp model.
For CTCs that are adequately described by the 1Comp model, the estimates of  one of the compartment volumes is close to zero. Like this and in contrast to a voxelwise approach parameter estimates are stable and easy to interpret.

Confronted with redundancy issues, modelling with compartments requires trade-off between too simplistic models and overparametrised, redundant models.
Complexity is determined by the number of compartments as well as the level of spatial resolution.
Spatial regularisation offers a solution that can be applied in other fields as well, for instance in the quantitative analysis of positron emission tomography (PET) and single-photon emission computed tomography (SPECT) images.
In PET and SPECT neuroreceptor imaging studies the kinetics of ligand uptake in the brain is described with the help of compartment models~\citep{Slifstein2001}. When estimating receptor parameters one copes with similar identifiability issues encountered in DCE-MRI analysis.

We have proposed $p_D$ as a measure that contains additional information about the heterogeneity of the tissue whereas the kinetic parameters contain information about the uptake dynamics only.
We find it interesting that estimates of the effective number of parameters $p_D$ rarely exceed values of 1.5, even for CTCs simulated from a 2Comp model with four kinetic parameters.

In summary, we proposed and evaluated spatial regularisation for two-com\-part\-ment models, allowing a more comprehensive insight into tissue perfusion, in particular in heterogeneous tissue. Spatial regularisation allows to overcome the redundancy issues by "borrowing strength" across the tissue of interest, and hence allows to fit complex compartment models even on voxel level with low signal-to-noise ratio. Additional clinical studies should be performed to further explore the clinical potential of this model.

\section*{Acknowledgements}
JS and VS are supported by Deutsche Forschungsgemeinschaft (DFG SCHM 2747/1-1).
Clinical data were graciously provided by Dr.~A.R.~Padhani, PSSC, Mount Vernon Hospital, Northwood, U.K.

\begin{appendix}
\section{APPENDIX}
\subsection{Likelihood and full conditionals}
\label{sec:appendixfc}
The log-likelihood depends on the voxel-specific kinetic parameters $\phi^{i}$
and the inverse noise variance $\tau_{\epsilon}=\frac{1}{\sigma^2}$:
\begin{eqnarray}
 l(\phi^{i},\tau_{\epsilon}) &=&\frac{T}{2} \log\left(2\pi\tau_{\epsilon}\right) - \frac{1}{2} \tau_{\epsilon} \sum^{T}_{j=1}\left(Y_{i,j} - C_{t}(\phi^{i}, t_j)\right)^2.
\end{eqnarray}

In the spatial model, the full conditional distribution of the logarithmic rate constant in voxel $i$, $\theta_{k}^i$, given the logarithmic rate constants of all other voxels, $\theta_{k}^{-i}$,  
$$
	p(\theta_{k}^i | \theta_{k}^{-i},\tau_{\theta_{k}}) \propto  \exp(-\frac{\tau_{\theta_{k}}}{2} \sum_{j \in \partial(i)} \left(\theta_{k}^i - \theta_{k}^j\right)^2)
$$
depends only on those of its direct neighbours for $k=1,2$. Here, $\partial(i)$ denotes the set of direct neighbours of voxel $i$. The full conditionals of the logarithmic transfer constants $\gamma_{k}^i$ have the same form.

Let $\epsilon_{ij} = Y_{i,j} - C_{t}(\phi^{i}, t_j)$ denote the random noise terms. 
Then, the full conditional of $\tau_{\epsilon}$   is
$\tau_{\epsilon} | \cdot \sim Ga(a+\frac{NT}{2}, b+\frac{1}{2}\sum^{N}_{i=1} \sum^{T}_{j=1} \epsilon_{ij}^2)$ for the spatial model.
The full conditional of the precision $\tau_{\theta_{1}}^i$ is
$\tau_{\theta_{1}}^i | \tau_{\theta_{1}}^{-i} \sim Ga(a_{\theta}+\frac{\left| \partial(i)\right|}{2}, b_{\theta}+\frac{1}{2}\sum_{j \in \partial(i)}\left(\tau_{\theta_{1}}^i-\tau_{\theta_{1}}^j\right)^2$.
Similarly for $\tau_{\theta_{2}}^i$, $\tau_{\gamma_{1}}^i$ and $\tau_{\gamma_{2}}^i$.

\end{appendix}

\bibliographystyle{chicago}
\bibliography{Literatur}

\end{document}